\documentclass[trackchanges,twocolumn]{aastex62}

\received{2019 February 4}
\revised{2019 March 2}
\accepted{2019 March 6}

\shorttitle{Clumpy Torus in Seyfert 1 Galaxies}
\shortauthors{Ogawa et al.}

\begin{document}

\title{Application of Clumpy Torus Model to Broadband X-ray Spectra of
Two Seyfert~1 Galaxies: IC~4329A and NGC~7469}


\author{Shoji Ogawa}
\affil{Department of Astronomy, Kyoto University, Kitashirakawa-Oiwake-cho, Sakyo-ku, Kyoto 606-8502, Japan}

\author{Yoshihiro Ueda}
\affiliation{Department of Astronomy, Kyoto University, Kitashirakawa-Oiwake-cho, Sakyo-ku, Kyoto 606-8502, Japan}

\author{Satoshi Yamada}
\affiliation{Department of Astronomy, Kyoto University, Kitashirakawa-Oiwake-cho, Sakyo-ku, Kyoto 606-8502, Japan}

\author{Atsushi Tanimoto}
\affiliation{Department of Astronomy, Kyoto University, Kitashirakawa-Oiwake-cho, Sakyo-ku, Kyoto 606-8502, Japan}

\author{Toshihiro Kawaguchi}
\affiliation{Department of Economics, Management and Information Science, Onomichi City University, Hiroshima 722-8506, Japan}

\begin{abstract}

We apply a new X-ray clumpy
torus model called XCLUMPY \citep{Tanimoto2019a},
where the clump distribution is assumed to be the same as in the
infrared clumpy torus model (CLUMPY) by \citet{Nenkova2008a,Nenkova2008b},
to the broadband X-ray spectra of type-1 active galactic nuclei
(AGNs) for the first time.
We analyze the archival data of IC~4329A and NGC~7469
observed with \textit{NuSTAR}/\textit{Suzaku} and \textit{NuSTAR}/\textit{XMM-Newton}, respectively, whose
infrared spectra were studied with CLUMPY by \citet{Ichikawa2015}
and optical extinctions ($A_{\rm V}$) of the tori were estimated.
We consider two models, invoking (Model 1) a relativistic reflection component
from the accretion disk and (Model 2) a partial absorber.
Assuming that the narrow Fe
K$\alpha$ emission line at 6.4~keV originates from the torus,
we separate the contribution of the torus reflection components in the
total spectra.
Our models yield equatorial hydrogen
column densities of the tori
to be $N^{\rm Equ}_{\rm H} =$ (0.53--1.43) $\times 10^{23}~\rm cm^{-2}$
and $N^{\rm Equ}_{\rm H} =$ (0.84--1.43) $\times 10^{24}~\rm cm^{-2}$,
for IC~4329A and NGC~7469,
respectively.
We find that the $N_{\rm H}/A_{\rm V}$ ratios in the tori are by factors of 25--68 (IC~4329A)
and 2.4--3.9 (NGC~7469) smaller than that in the Galactic interstellar medium (ISM).
These results suggest that a non-negligible fraction of AGNs
are ``dust-rich'' compared with the Galactic ISM, as
opposite to the general trend previously reported in many obscured AGNs.

\end{abstract}

\keywords{
X-rays: galaxies --
galaxies: Seyfert --
galaxies: individual (\objectname{IC~4329A, NGC~7469})
}

\section{Introduction}
\label{sec1}

To reveal basic properties of obscuring material
in active galactic nuclei (AGNs), often referred as the ``torus'',
is important to understand feeding and feedback mechanisms of AGNs
\citep[see e.g.,][for a recent review]{RamosAlmeidaRicci2017}. Among
them, the gas-to-dust ratio is a key parameter to understand
the circumnuclear environments. It has been
reported in many (but not all) AGNs that the ratios of optical
extinction ($A_{\rm V}$) to hydrogen column density ($N_{\rm H}$) toward the nuclei, as
estimated from the infrared/optical and X-ray spectra, respectively, are
smaller than the Galactic value \citep{Maiolino2001,Vasudevan2009,Burtscher2016}.
A plausible explanation is that the gas-to-dust ratio of obscuring
material is higher  (i.e., more ``gas rich'') than that of the Galactic
interstellar medium (ISM). It may be due to dust-free neutral gas in the
broad line region (BLR), which can also cause variability in
the X-ray absorption (see \citealt{Burtscher2016} and references therein).
However, opposite cases (i.e., torus is more ``dust-rich'' than
Galactic) have also been reported \citep[e.g.,][]{Barcons2003,Huang2011,Ordovas-Pascual2017},
making our understanding of AGN
environments not that simple. More independent studies using a well
studied, local AGN sample are necessary to solve this issue.

The X-ray spectrum of an AGN contains a reflection component from the
torus, accompanied by narrow fluorescence lines such as Fe K$\alpha$
at 6.4~keV. This component carries information of all material {\it
including gas and dust} around the supermassive black holes (SMBH). In
particular, even in type-1 AGNs where no line-of-sight absorption is
observed, the equivalent width of the Fe K$\alpha$ line can be used to
infer the torus structure, such as its covering fraction and/or column
density  \citep[e.g.,][]{Tazaki2013,Kawamuro2016LLAGN}. Complementary to
the X-ray data, the infrared spectra give information on the properties
of {\it dust}. Thus, comparing the X-ray and infrared spectra is quite
useful to constrain the nature of AGN tori \citep[e.g.,][]{Ricci2014A&A},
including the gas-to-dust ratio. In such
studies, it is desired to apply ``self-consistent'' models in terms of the
torus geometry to both X-ray and infrared data.

Recently, \citet{Tanimoto2019a} have constructed a new X-ray clumpy torus
model called XCLUMPY, based on the Monte Carlo simulation for
Astrophysics and Cosmology (MONACO: \citealt{Odaka2016}) framework.  In
this model, the geometry of the torus is the same as that in the CLUMPY
model in the infrared band \citep{Nenkova2008a,Nenkova2008b}, where
clumps are distributed according to power law and normal profiles in the
radial and angular directions, respectively. It has three variable torus
parameters: equatorial hydrogen column density, torus angular width, and
inclination angle. The XCLUMPY model enables us to directly compare the
results with those obtained from the infrared spectra with the CLUMPY
code in a self-consistent way.

In this paper, we apply the XCLUMPY model to the X-ray spectra of
unobscured (type-1) AGNs for the first time. Our sample is IC~4329A and
NGC~7469, whose infrared spectra have been analyzed in detail with the
CLUMPY model \citep{Ichikawa2015}. We utilize their best-quality broadband
X-ray spectra, simultaneously observed with \textit{NuSTAR} and \textit{Suzaku} (for IC~4329A)
and with \textit{NuSTAR} and \textit{XMM-Newton} (NGC~7469). This work is
complementary to those for obscured AGNs that compare the line-of-sight
column densities with the $A_{\rm V}$ values obtained with the CLUMPY model
(\citealt{Gonzalez2013}; Tanimoto et al., in prep.).
The main goal of our paper is to constrain the torus
properties, in particular the gas-to-dust ratios. In addition, we can
also correctly estimate the contribution of the reflection component
from the torus in the total spectra.
Two major models have been proposed as the X-ray spectra of type-1 AGNs
to explain the broad iron-K emission line feature
and bump structure peaked around 30 keV:
one invoking a relativistic reflection component from the innermost
accretion disk, the other assuming variable partial absorbers (see
Section~3). Although to discriminate these models is not a purpose of
our paper (and hence we treat the two models equally), it is always
crucial to properly take into account the torus reflection component in
modeling the broadband X-ray spectra of AGNs.

This paper is organized as follows.
Section 2 gives the details of our sample.
In Section 3, we
describe the observations and data reduction. In Section 4, we
present the analysis of the broadband X-ray spectra by applying the
XCLUMPY model. In Section 5, we compare our
results with the previous studies and discuss their torus properties.
We adopt
the cosmological parameters of $(H_{0}, \Omega _{m}, \Omega_{\Lambda})
= (70~\rm{km~s^{-1}~M~pc^{-1}}, 0.3, 0.7)$ and
the solar abundances of \citet{Anders&Grevesse1989} throughout the
paper. Errors on spectral parameters correspond to 90\% confidence
limits for single parameters.

\section{Sample}
\label{sec2}

For our study we selected two Seyfert~1 galaxies, IC~4329A ($z=0.0161$;
\citealt{Willmer1991}) and NGC~7469 ($z=0.0163$;
\citealt{Springob2005}), from the sample of \citet{Ichikawa2015}. To
constrain the torus parameters, \citet{Ichikawa2015} applied the CLUMPY
model to the infrared spectral energy distributions of 21 nearby AGNs obtained with
\textit{Spitzer} and high-spatial resolution cameras on ground-based
telescopes. Among them, these are the two AGNs that show low X-ray
absorptions ($N_{\rm H} < 10^{23}~\rm cm^{-2} $). We excluded NGC~4151 and NGC~1365,
which are classified as Seyfert~1 galaxies but are known to show very
complex, variable absorption in the X-ray bands (see e.g., \citealt{Yaqoob1991} and \citealt{Risaliti2005},
respectively).

The X-ray data of these sources were analyzed by many authors (e.g., for IC~4329A,
\citealt{Done2000} (\textit{ASCA}+\textit{RXTE}),
\citealt{Gondoin2001} (\textit{XMM-Newton}+\textit{BeppoSAX}),
\citealt{McKernan2004} (\textit{Chandra}/HETGS), \citealt{Steenbrugge2005}
(\textit{XMM-Newton}),
\citealt{Beckmann2006} (\textit{INTEGRAL}),
\citealt{Winter2009} (\textit{Swift}/BAT),
\citealt{Patrick2012} (\textit{Suzaku}+\textit{Swift}/BAT),
\citealt{Miyake2016} (\textit{Suzaku}),
\citealt{Iso2016} (\textit{Suzaku}) and \citealt{Brenneman2014}
(\textit{Suzaku}+\textit{NuSTAR});
for NGC~7469, e.g.,
\citealt{Guainazzi1994} (\textit{ASCA}), \citealt{Nandra2000}
(\textit{RXTE}) \citealt{Rosa2002} (\textit{BeppoSAX}),
\citealt{Blustin2003} (\textit{XMM-Newton}),
\citealt{Scott2005} (\textit{Chandra}/HETGS),
\citealt{Winter2009} (\textit{Swift}/BAT),
\citealt{Patrick2012} (\textit{Suzaku}+\textit{Swift}/BAT),
\citealt{Iso2016}
(\textit{Suzaku}) and \citealt{Middei2018} (\textit{XMM-Newton}+\textit{NuSTAR})).
The \textit{Chandra}/HETGS observations \citep{Shu2010}
confirmed the presence of a narrow
($<10,000$ km s$^{-1}$ in FWHM) Fe K$\alpha$ component
centered at 6.4~keV in both IC~4329A (see their Appendix) and NGC~7469;
in IC~4329A, excess emission is detected at $\approx$6.0--6.4~keV, which can be interpreted as a modestly broadened Fe K$\alpha$ line from
the disk \citep{Brenneman2014}.
Many of these works employed
the {\tt pexrav} model \citep{Magdziarz1995} with an Fe K$\alpha$ line,
the {\tt pexmon} model \citep{Nandra2007}, which includes self-consistently calculated
fluorescence lines, or the {\tt xillver} model \citep{Garcia2014} whose
ionization parameter is fixed at zero, to represent the reflection
component from the torus, or to approximate
total reflection components including that from the accretion disk.
Models of relativistic reflection from the accretion disk
have been applied by \citet{Done2000} and \citet{Patrick2012} for IC~4329A,
and by \citet{Rosa2002} and \citet{Patrick2012} for NGC~7469.
\citet{Iso2016} systematically applied a
partial covering model
to local Seyfert galaxies including our targets.

\begin{deluxetable*}{llcllc}
\tablewidth{\textwidth}

\tablecaption{Summary of Observations \label{tab1-obs}}
\tablehead{
Object         &
Satellite      &
ObsID          &
Start Date(UT) &
End Date(UT)   &
Net Exposure(ks)
}
\startdata
IC~4329A    & \textit{Suzaku}       & 707025010     & 2012-08-13 02:13:09 & 2012-08-14 10:53:03 & 117.6 \\
            & \textit{NuSTAR}       & 60001045002   & 2012-08-12 16:06:07 & 2012-08-14 13:12:46 & 162.4 \\
NGC~7469    & \textit{XMM-Newton}   & 0760350201    & 2015-06-12 13:36:49 & 2015-06-13 14:50:09 & 90.8 \\
            & \textit{NuSTAR}       & 60101001002   & 2015-06-12 18:41:07 & 2015-06-13 00:40:46 & 21.6
\enddata

\end{deluxetable*}

\section{Observations and Data Reduction}
\label{sec3}

We analyze the best-quality simultaneous broadband X-ray spectra that
cover the energy band from 0.3~keV to 70~keV with a CCD energy resolution
below $\sim$10~keV. IC~4329A was observed simultaneously with
\textit{Suzaku} \citep{Mitsuda2007} and \textit{NuSTAR}
\citep{Harrison2013} in 2012, and NGC~7469 was with \textit{XMM-Newton}
\citep{Jansen2001} and \textit{NuSTAR} in 2015. The observation log of
the X-ray data used in this paper is given in Table
\ref{tab1-obs}. Details of data reduction are described below.

\subsection{IC~4329A}

\subsubsection{\textit{Suzaku}}

\textit{Suzaku}
observed IC~4329A in 2012 August. It carried four X-ray CCD
cameras (X-ray imaging spectrometer; XISs \citep{Koyama2007})
and a non-imaging instrument (the hard X-ray detector; HXD \citep{Takahashi2007}),
which cover the
energy band below and above $\approx$10~keV, respectively. XIS0, XIS2, and
XIS3 are frontside-illuminated CCDs (XIS-FI) and XIS1 is the
backside-illuminated one (XIS-BI). The HXD consists of the PIN (10--70~keV) and GSO (40--600~keV) detectors \citep{Kokubun2007}.

We reprocessed the unfiltered XIS event data with the \textsc{
aepipeline} script. Events were extracted from a circular region with a
radius of 160 arcsec centered at the source position. The background was
taken from a source-free circular region with a radius of 120 arcsec. We
generated the response matrix file (RMF) with \textsc{ xisrmfgen}
and ancillary response files (ARF) with \textsc{ xissimarfgen}
\citep{Ishisaki2007}. To improve the statistics, we co-added the spectra
of XIS0 and XIS3. The spectrum were binned to contain at least 1000
counts per bin. We did not utilize the data of XIS1, whose effective
area in the iron-K band was smaller than those of XIS-FIs, to avoid
cross-calibration uncertainties.

The unfiltered HXD data were also reprocessed by using \textsc{
aepipeline}. We only analyzed the PIN data, because the source was
weakly detected with the GSO \citep{Brenneman2014}.
We utilized the ``tuned'' background event files \citep{Fukazawa2009} to
make the spectrum of the non-X-ray background (NXB), to which a
simulated spectrum of the cosmic X-ray background was added.
In the spectral analysis, we only utilized the 16--40~keV range, where
the source flux is brighter than 3\% of the NXB level (the maximum
systematic error in the 15--70~keV range; see \citealt{Fukazawa2009}).
\\
\subsubsection{\textit{NuSTAR}}

\textit{NuSTAR} also observed IC~4329A in 2012 August simultaneously
with the \textit{Suzaku} observation. \textit{NuSTAR} carries two forcal
plane modules (FPMs: FPMA and FPMB), which cover an energy range of
3--79~keV. We analyzed the FPMs data with HEAsoft v6.21 and CALDB
released in 2017 December 12. We extracted the spectrum from a circular
region with a 75 arcsec radius centered at the source peak, and took the
background from a nearby source-free circular region with the same
radius. We then combined the source spectra, background spectra, RMF,
and ARF, using the \textsc{ addascaspec} script.  The combined spectrum
was binned to contain at least 1000 counts per bin.

\subsection{NGC~7469}

\subsubsection{\textit{XMM-Newton}}

\textit{XMM-Newton} observed NGC~7469 in 2015 June. It carries three
X-ray CCD cameras, one EPIC/PN \citep{Struder2001} and MOS \citep{Turner2001}.
We analyzed only the data of PN, which has much larger
effective area than the MOS detectors, using the Science Analysis Software
(SAS) v17.0.0 and current calibration file (CCF) released on 2018 June 22.
The PN data were reprocessed with the \textsc{ epproc} script. We
extracted the spectrum from a circular region with a radius of 40 arcsec
centered at the source peak, and took the background from a source-free
circular region with a 50 arcsec radius in the same CCD chip. We
generated the RMF with \textsc{ rmfgen} and ARF with \textsc{
arfgen}. The spectrum was binned to contain at least 100 counts per bin.

\subsubsection{\textit{NuSTAR}}

NGC~7469 was also observed with \textit{NuSTAR} in 2015 June
simultaneously with \textit{XMM-Newton}. We extracted the FPMs
spectra from a circular region with a 70 arcsec radius centered at the
source peak, and took the background from a nearby source-free circular
region with the same radius. We combined the spectra of FPMs, using \textsc{ addascaspec}.
The combined spectrum was binned to contain at least 100 counts per bin.

\begin{deluxetable*}{llllll}
\tablewidth{\textwidth}

\tablecaption{Best-fit Parameters of IC~4329A \label{tab2-par}}
\tablehead{
Component      &
No.            &
Parameter      &
Model 1       &
Model 2        &
Units
}
\startdata
{\tt ZXIPCF1}  & (1) & $N_{\rm H}$ & $5.68^{+0.06}_{-0.12} $ & $5.92^{+0.13}_{-0.12} $ & $ \rm 10^{21} \, cm^{-2} $ \\
                & (2) & $\log \xi $ & $0.27^{+0.01}_{-0.10} $ & $0.17^{+0.10}_{-0.09} $ &  \\
{\tt ZXIPCF2}  & (3) & $N_{\rm H}$ & \nodata & $1.04\pm0.06 $ & $ \rm 10^{24} \, cm^{-2} $ \\
                & (4) & $\log \xi $ & \nodata & $1.91^{+0.06}_{-0.14} $ &  \\
                & (5) & $C_{\rm frac}$ & \nodata & $0.23^{+0.01}_{-0.02} $ & \\
\hline
{\tt ZCUTOFFPL} & (6) & $\Gamma$ & $1.77\pm0.01 $ & $1.83\pm0.01 $ & \\
                & (7) & $E_{\rm Cut} $ & $318^{+53}_{-60}$ & $ 1000^{+0}_{-115}\tablenotemark{a} $ & $ \rm keV $ \\
                & (8) & $K_{\rm P} $ & $3.05^{+0.02}_{-0.03}  $ & $4.21^{+0.13}_{-0.14} $ & $10^{-2}\, \rm photon\,cm^{-2}\,s^{-1}$ \\
\hline
{\tt ZGAUSS}    & (9) & $ E_{\rm Line}$        & $0.77\pm0.01$              & $0.77\pm0.01$ & $\rm keV $ \\
                & (10) & $\sigma_{\rm Line} $      & $2.98^{+0.27}_{-0.55} $                      & $ 3.17\pm0.58 $& $ 10^{-2} \, \rm keV $ \\
                & (11) & $K_{\rm L} $    & $4.07^{+0.37}_{-0.44}$ & $3.99^{+0.51}_{-0.46} $ & $  10^{-4} \, \rm photon\,cm^{-2}\,s^{-1}$ \\
\hline
{\tt RELXILL}   & (12) & $\log \xi $ & $ 0.29^{+0.71}_{-0.29} $ & \nodata &  \\
                & (13) & $R_{\rm in} $ & $87^{+73}_{-31}$ & \nodata & $  r_{\rm G} $ \\
                & (14) & $ R $ & $ 3.20^{+0.13}_{-0.25} $ & \nodata & $10^{-3}$ \\
\hline
{\tt XCLUMPY}   & (15) & $N^{\rm Equ}_{\rm H}$ & $ 0.66^{+0.11}_{-0.13} $ & $1.35^{+0.08}_{-0.09}  $ & $ \rm 10^{23} \, cm^{-2} $ \\
                & (16) & $\sigma $ & $40$ (fixed) & $40$ (fixed) & $\rm degree$ \\
                & (17) & $ i $\tablenotemark{b} & $18.19$ (fixed) & $18.19$ (fixed) & $\rm degree$  \\
\hline
                & (18) & $C_{\textit{NuSTAR}}$ &     $0.97\pm0.01 $       &     $0.97\pm0.01 $       & \\
                & (19) & $L_{\rm 2-10} $    & $6.47$ & $8.14 $ & $  10^{43} \, \rm erg\,s^{-1}$ \\
                &  & $\chi^2$/dof &     $1338.8/1120$      &      $1321.2/1120$      &
\enddata
\tablecomments
{
(1) Hydrogen column density of a full absorber.
(2) Its logarithmic ionization parameter, $\xi$ ($\rm erg~cm~s^{-1}$).
(3) Hydrogen column density of a partial absorber.
(4) Its logarithmic ionization parameter.
(5) Its covering fraction.
(6) Photon index.
(7) Cutoff energy.
(8) Power-law normalization of the direct component.
(9) Energy of the emission line.
(10) Line width of the emission line.
(11) Normalization of the emission line.
(12) Logarithmic ionization parameter of the accretion disk.
(13) The inner radius of the disk.
(14) Reflection strength.
(15) Torus hydrogen column density along the equatorial plane.
(16) Torus angular width.
(17) Inclination angle.
(18) Cross-calibration constant of \textit{NuSTAR} relative to \textit{Suzaku}/XIS.
(19) Intrinsic luminosity in the 2--10~keV.
}
\tablenotetext{a}{The upper limit is pegged at the upper boundary value in the table model.}
\tablenotetext{b}{Because the infrared result ($i = 4^{+4}_{-3}$~degrees) is out
of the allowed range of XCLUMPY, we fix it at its lower limit
 ($18.19$~degrees); differences from $i=4$~degrees are expected to be negligible \citep{Tanimoto2019a}. }
\end{deluxetable*}

\begin{deluxetable*}{llllll}
\tablewidth{\textwidth}

\tablecaption{Best-fit parameters of NGC~7469\label{tab3-par}}
\tablehead{
Component      &
No.            &
Parameter      &
Model 1        &
Model 2        &
Units
}
\startdata
{\tt ZXIPCF1}  & (1) & $N_{\rm H} $ & $1.30^{+0.15}_{-0.20} $ & $1.56^{+0.13}_{-0.14} $  & $ \rm 10^{21} \, cm^{-2} $ \\
                & (2) & $\log \xi $ & $2.49\pm0.06 $ & $2.40\pm0.05 $  &  \\
{\tt ZXIPCF2}  & (3) & $N_{\rm H} $ & \nodata & $1.80^{+0.22}_{-0.21} $ & $ \rm 10^{24} \, cm^{-2} $ \\
                & (4) & $\log \xi $ & \nodata & $3.18^{+0.22}_{-0.14} $ &  \\
                & (5) & $C_{\rm frac}$ & \nodata & $0.15\pm0.04 $ & \\
\hline
{\tt ZCUTOFFPL} & (6) & $\Gamma$        & $1.92^{+0.06}_{-0.04}$              & $1.84^{+0.04}_{-0.02}$ & \\
                & (7) & $E_{\rm Cut} $  & $322^{+678}_{-122}\tablenotemark{a} $  & $ 321^{+679}_{-139}\tablenotemark{a} $& $ \rm keV $ \\
                & (8) & $K_{\rm P} $    & $9.22^{+0.96}_{-0.60}$ & $8.90^{+1.04}_{-0.63} $ & $  10^{-3} \, \rm photon\,cm^{-2}\,s^{-1}$ \\
\hline
{\tt COMPTT}    & (9) & $T_{\rm bb}$        & $8.10^{+0.74}_{-0.35}$              & $8.61^{+0.18}_{-0.12}$ & $ 10^{-2} \, \rm keV $ \\
                & (10) & $T_{\rm P} $ & $2.40^{+1.71}_{-2.40} $ & $ 9.60^{+2.28}_{-1.09} $& $ \rm keV $ \\
                & (11) & $ \tau $    & $3.21^{+0.56}_{-0.46}$ & $1.04^{+0.18}_{-0.14} $ &  \\
                & (12) & $K_{\rm S} $    & $1.68^{+0.57}_{-0.81}$ & $5.11^{+9.65}_{-0.81} $ & $  10^{-2} \, \rm photon\,cm^{-2}\,s^{-1}$ \\
\hline
{\tt RELXILL}   & (13) & $\log \xi $  & $0.99^{+0.75}_{-0.26} $ & \nodata &  \\
                & (14) & $R_{\rm in} $     & $9.4^{+26}_{-3.4} $ & \nodata  & $  r_{\rm G} $ \\
                & (15) & $ R $       & $2.64^{+0.23}_{-0.50}  $ & \nodata  & $10^{-3}$ \\
\hline
{\tt XCLUMPY}   & (16) & $N^{\rm Equ}_{\rm H} $    & $1.00^{+0.32}_{-0.17} $ & $1.00 ^{+0.43}_{-0.16} $ & $ \rm 10^{24} \, cm^{-2} $ \\
                & (17) & $ \sigma $   & $21$ (fixed)                 & $21$ (fixed) & $\rm degree$ \\
                & (18) & $ i $       & $59$  (fixed)                & $59$ (fixed) & $\rm degree$ \\
\hline
                & (19) & $C_{\textit{NuSTAR}}$ &     $1.13\pm0.01$      &     $1.13\pm0.01 $       & \\
                & (20) & $L_{\rm 2-10} $    & $1.68$ & $1.83 $ & $  10^{43} \, \rm erg\,s^{-1}$ \\
                &       & $\chi^2$/dof &     $1668.7/1576$      &      $1683.6/1576$      &
\enddata
\tablecomments
{
(1) Hydrogen column density of a full absorber.
(2) Its logarithmic ionization parameter, $\xi$ ($\rm erg~cm~s^{-1}$).
(3) Hydrogen column density of a partial absorber.
(4) Its logarithmic ionization parameter.
(5) Its covering fraction.
(6) Photon index.
(7) Cutoff energy.
(8) Power-law normalization of the direct component.
(9) Input soft photon temperature.
(10) Plasma temperature.
(11) Plasma optical depth.
(12) Normalization.
(13) Logarithmic ionization parameter of the accretion disk.
(14) The inner radius of the disk.
(15) Reflection strength.
(16) Torus hydrogen column density along the equatorial plane.
(17) Torus angular width.
(18) Inclination angle.
(19) Cross-calibration constant of \textit{NuSTAR} relative to \textit{XMM-Newton}/EPIC-PN.
(20) Intrinsic luminosity in the 2--10~keV.
}
\tablenotetext{a}{The upper limit is pegged at the upper boundary value in the table model.}
\end{deluxetable*}

\begin{figure*}
\plottwo{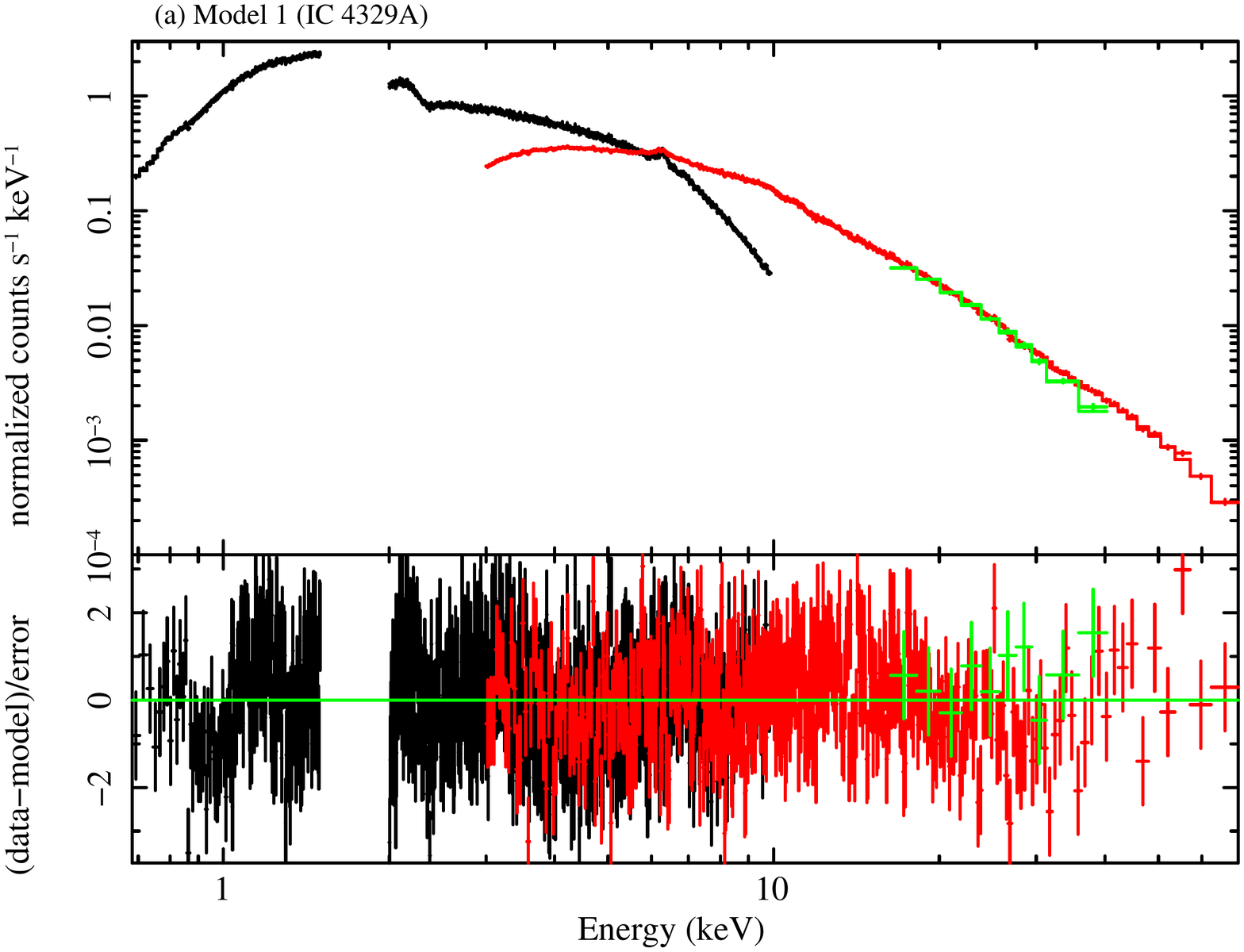}{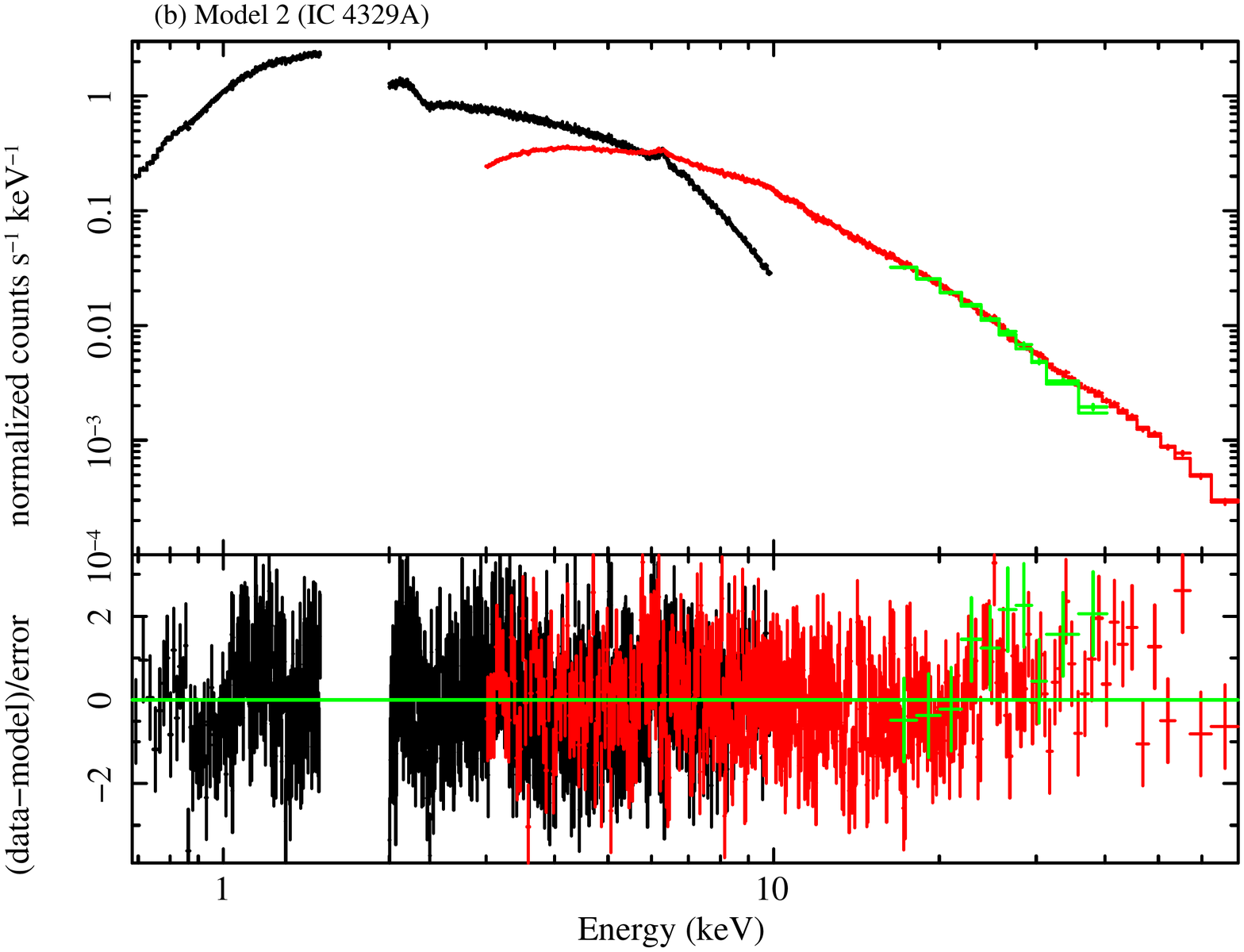}
\plottwo{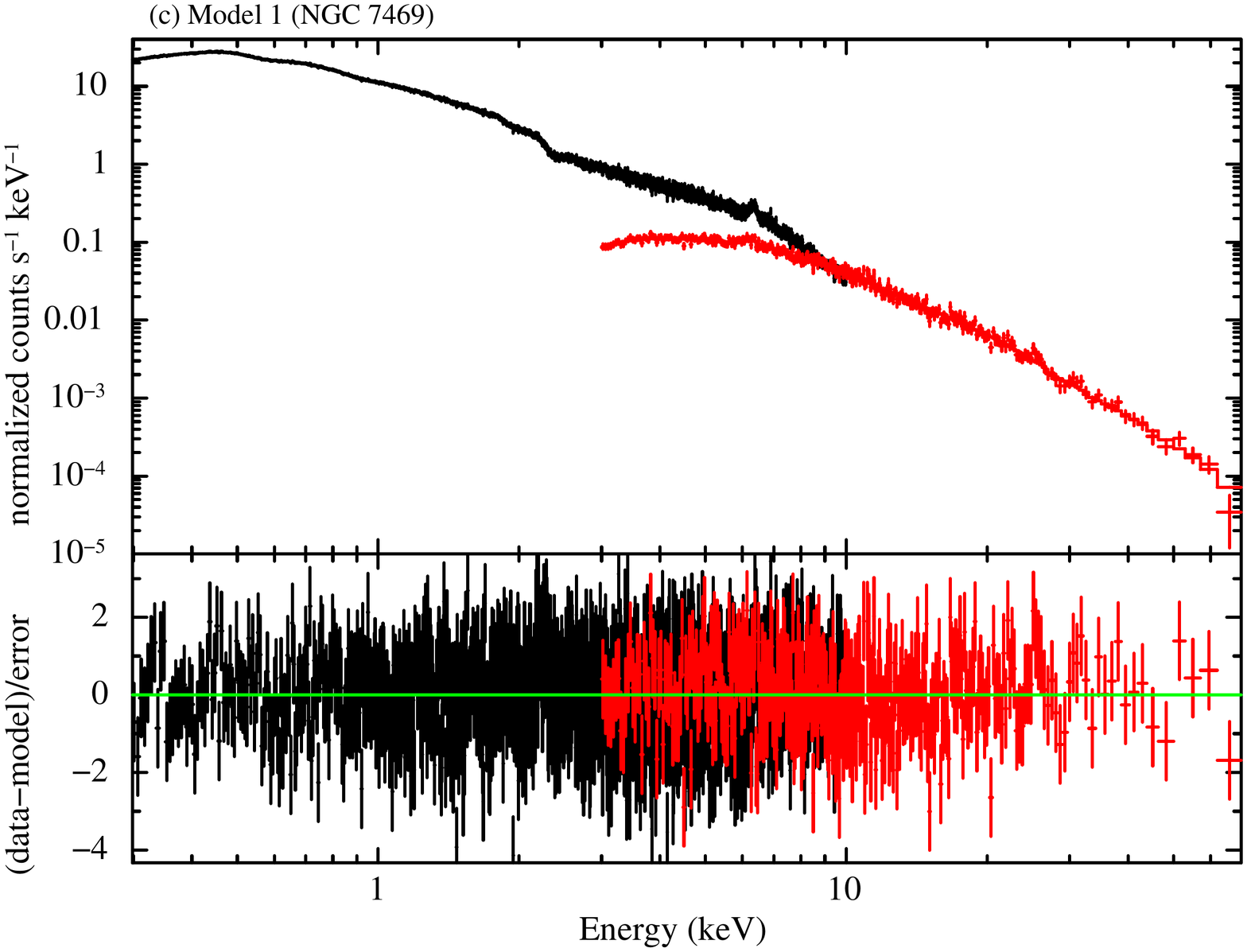}{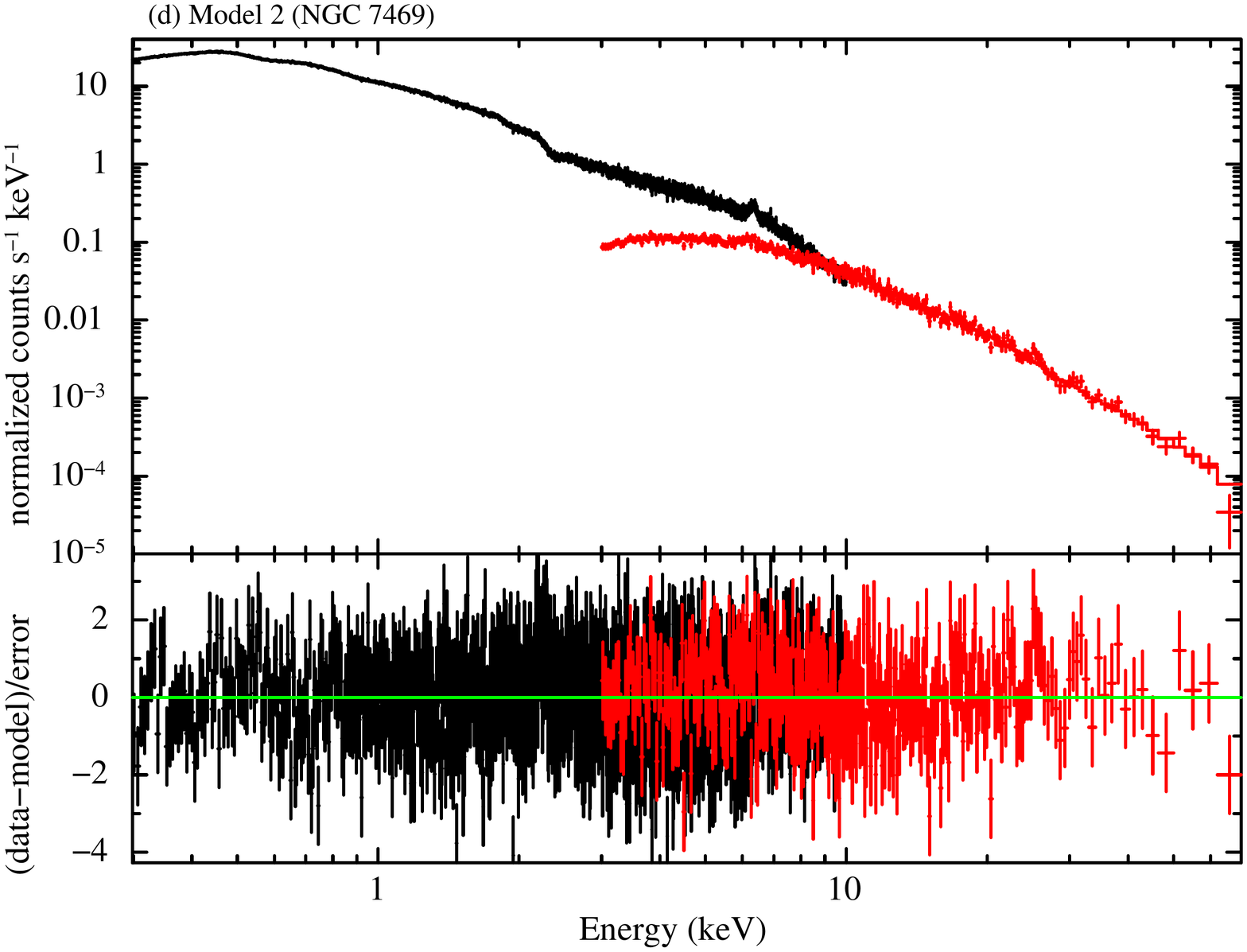}
\caption{
Observed broadband spectra of IC~4329A and NGC~7469 folded with the energy
responses. The best-fit models are overplotted. Left: Model 1. Right:
Model 2. upper: IC~4329A. Lower: NGC~7469. In the upper panels,
the folded spectra of \textit{Suzaku}/XIS (black crosses),
\textit{Suzaku}/HXD-PIN (green crosses), and \textit{NuSTAR}/FPMs (red
crosses) are plotted for IC~4329A,
and those of \textit{XMM-Newton}/EPIC-PN (black crosses) and
\textit{NuSTAR}/FPMs (red crosses) are plotted for NGC~7469. Solid curves
represent the best-fit models. In the lower panels, the fitting
residuals in units of 1$\sigma$ error are shown.
}
\label{fig1-fitting}
\end{figure*}

\begin{figure*}
\plottwo{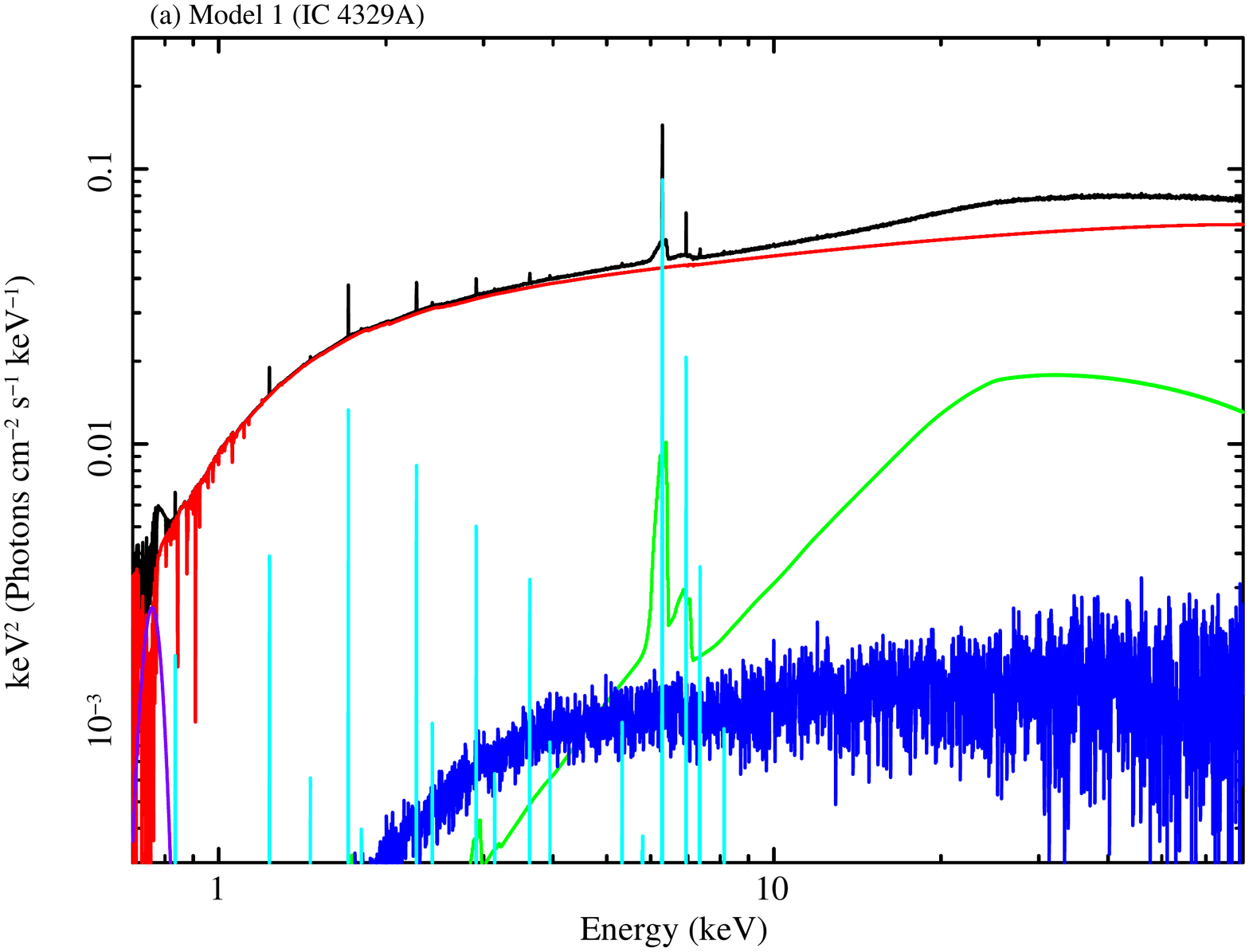}{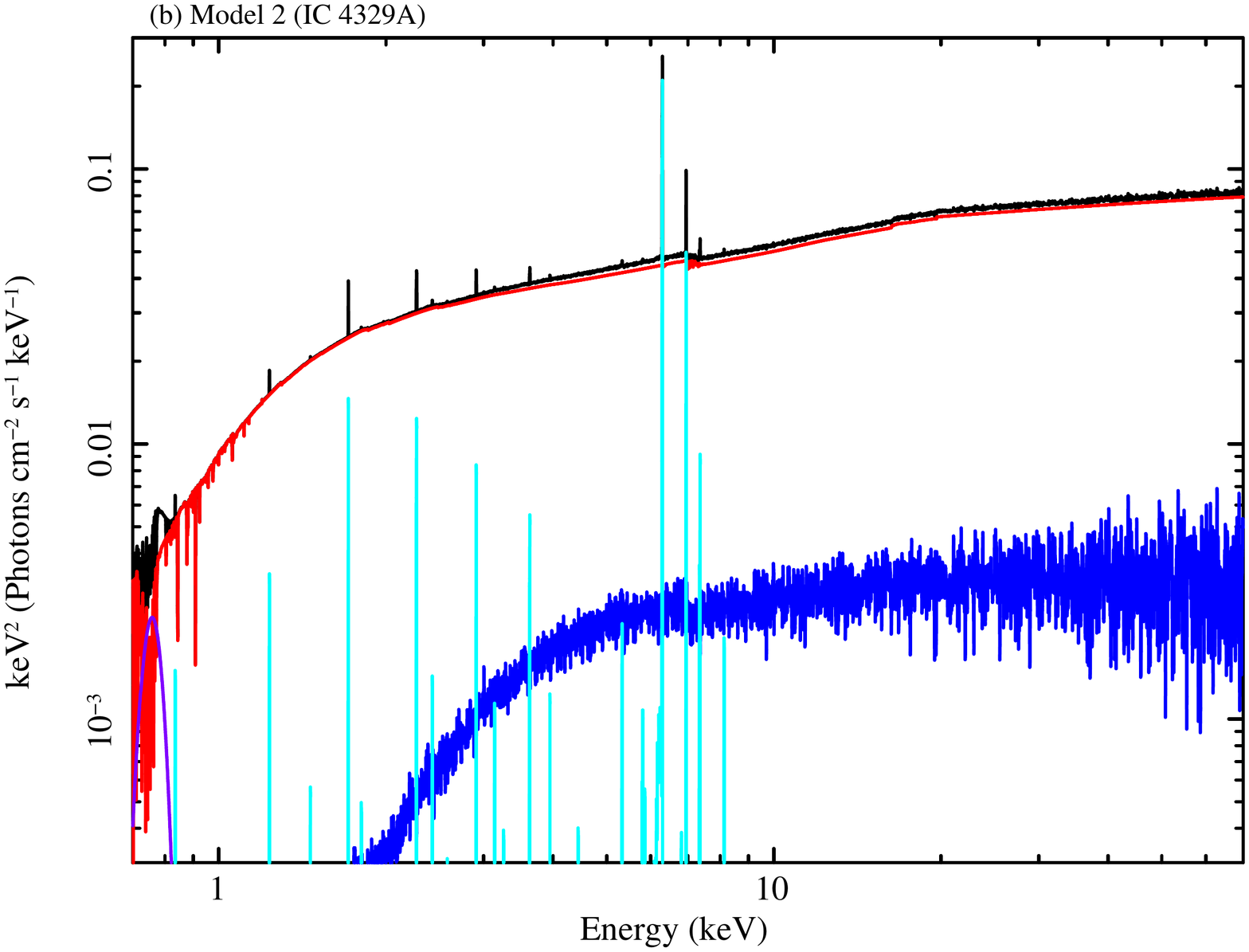}
\plottwo{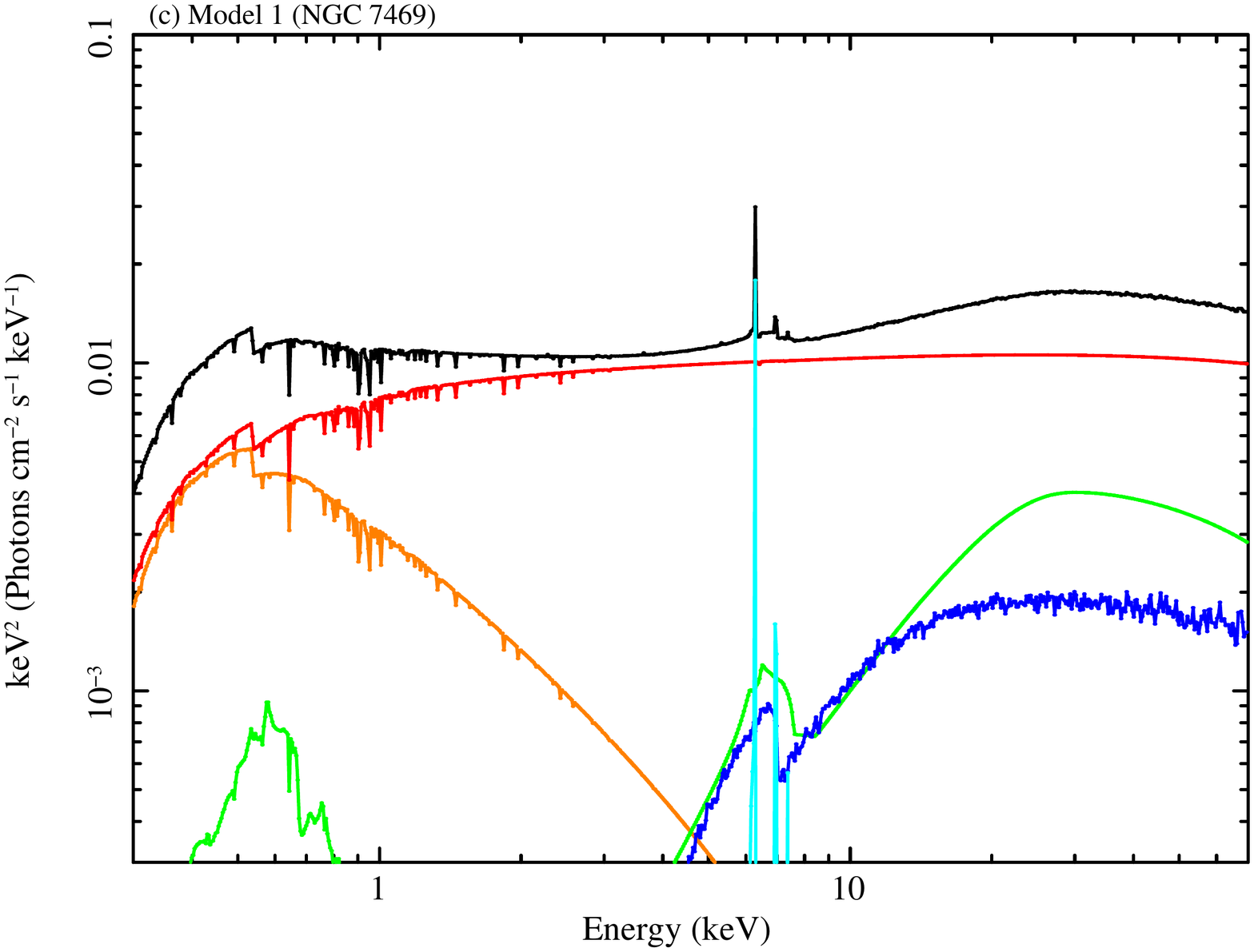}{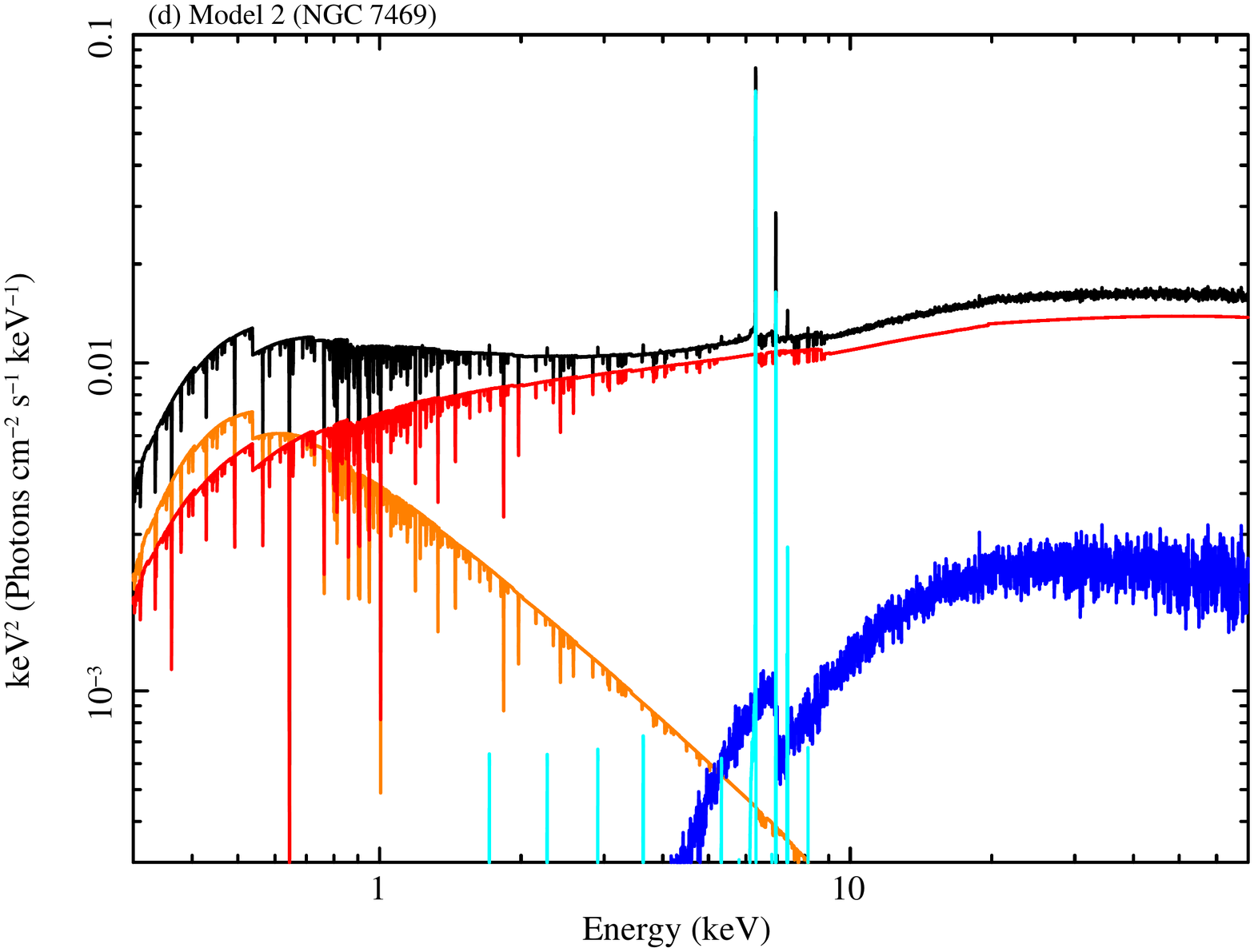}
\caption{
The best-fit models in units of $E I_E$ (where $I_E$ is the energy flux
 at the energy $E$). Left: Model 1. Right: Model 2. upper: IC~4329A.
 Lower: NGC~7469.Black line: total. Red line: direct component. Green line: reflection component from the
 accretion disk. Blue line: reflection continuum from the torus. Light blue line: emission
lines from the torus. Orange line: soft excess. Purple line: emission line at 0.77~keV.  \\
}

\label{fig2-models}
\end{figure*}

\section{X-ray Spectral Analysis and Results}
\label{sec4}

It is well established from previous studies that a typical X-ray
spectrum of Seyfert~1 galaxies cannot be represented by a single power
law but consists of multiple components: a direct power-law component
with an exponential cutoff (or thermally Comptonized component), its
reflection components from the torus and/or the accretion disk, and an
soft excess \citep[see e.g.,][]{Risaliti2004}.
Absorptions by ionized matter (warm
absorbers) are often observed. In the iron-K
band, in addition to a narrow Fe K$\alpha$ emission line centered at 6.4~keV,
a broad emission line feature is sometimes recognized, although its
intensity and shape strongly depend on the continuum modelling. The
hard X-ray continuum shows a bump structure peaked around 30~keV over a
power law component extrapolated from lower energies. The
origins of these spectral features are still in debate, for which at
least two distinct interpretations\footnote{Another interpretation
considering dual power-law components have been proposed by e.g.,
\citet{Noda2013a}, see also \citet{Kawaguchi2001} for an earlier theoretical work.}
have been proposed: (1) relativistically blurred reflection
from the accretion disk \citep[e.g.,][]{Tanaka1995}, and (2) partial covering by
line-of-sight absorbers with multiple ionization stages \citep[e.g.,][]{Miyakawa2012}.

We perform simultaneous fit to the
\textit{Suzaku}/XIS (0.7--10~keV), \textit{Suzaku}/HXD-PIN (16--40~keV), and
\textit{NuSTAR}/FPMs (3--70~keV) data for IC~4329A, and
\textit{XMM-Newton}/EPIC-PN (0.3--10~keV) and \textit{NuSTAR}/FPMs (3--70~keV)
for NGC~7469. We ignore the 1.5--2.0 keV range for IC 4329A
to avoid possible calibration uncertainties in the \textit{Suzaku}/XIS spectra.

The observed spectra folded with the energy responses are
plotted in the upper panels of Figure~\ref{fig1-fitting}(a)-(d).

Following major previous works, here we consider two spectral models:
Model~1 where a relativistic reflection component from the accretion
disk is included, for which we employ the RELXILL model \citep{Dauser2013,Garcia2014}, and
Model~2 where partial covering is applied to the direct component, for
which we adopt the same model used by \citet{Iso2016}. We consider warm
absorber of one layer (Model~1) or two layers (Model~2). In both models,
we utilize the XCLUMPY model to represent the reflection component (with
fluorescence emission lines) from the torus. Model~1 and 2 are
commonly applied to the spectra of both targets.

We always consider Galactic absorption, whose column density is fixed at
the value estimated from the H~I map by \citep{Kalberla2005} for each
target. To correct for relative calibration differences in the effective
area among the instruments, we multiply a constant factor to the
spectra.  For IC~4329A, it is fixed at unity for \textit{Suzaku}/XIS and at 1.16
for \textit{Suzaku}/HXD-PIN (based on the calibration with the Crab Nebula), and is
left free for \textit{NuSTAR}/FPMs. For NGC~7469, it is fixed at unity for
\textit{XMM-Newton}/EPIC-PN and is left free for \textit{NuSTAR}/FPMs.

\subsection{Model 1: Relativistic Reflection Model}

Model 1 is composed of a direct power-law component, its relativistic
reflection component from the inner accretion disk, and that from the
torus. The first two components are subject to a warm absorber in the
line-of-sight. In the XSPEC terminology, the model is expressed as:

\begin{eqnarray}
    \textrm{Model 1} & = & \mathsf{const * phabs} \nonumber\\
    & * &\mathsf{(zxipcf *(zcutoffpl + compTT + relxill)} \nonumber\\
    & + &\mathsf{ atable\{xclumpy\_R.fits\} + atable\{xclumpy\_L.fits\}} \nonumber\\
    & + &\mathsf{ zgauss}). \nonumber\\
\end{eqnarray}

\begin{enumerate}
\renewcommand{\labelenumi}{(\arabic{enumi})}
\item The {\tt const} and {\tt phabs} terms represent the cross-calibration
constant and the Galactic absorption, respectively.

\item The {\tt zcutoffpl} term represents the direct component (cutoff
power-law), the {\tt compTT} term the soft excess (thermal
Comptonization model by \citealt{Titarchuk1994}), and the {\tt relxill}
term the reflection component from the
accretion disc based on the RELXILL code. All components are subject to
absorption by a warm absorber ({\tt zxipcf}).
The {\tt compTT} term is required only in NGC~7469.

RELXILL is a state-of-art relativistic reflection model,
which combines the XILLVER code \citep{Garcia2014}, a reflection
model from an ionized disk, with relativistic broadening by the RELLINE
code \citep{Dauser2010,Dauser2013}.
The free parameters of RELXILL
are the inclination angle of the observer with respect to the accretion
disk $(i)$, the iron abundance $(Z_{\rm Fe})$, the photon
index, the ionization parameter of the disk $(\xi)$, the fraction of
reflected flux ($R = \Omega / 2\pi $, where $\Omega$ is the solid angle
of the reflector), the spin parameter of the SMBH $(a)$, the inner $(q_1)$ and
outer $(q_2)$ emissivity indices, the radius at which the
emissivity index changes $(R_{\rm br})$, the inner $(R_{\rm in})$ and
outer $(R_{\rm out})$ radi of the disk, and the cutoff energy.
We assume $Z_{\rm Fe} = 1.0 $ (solar abundance), $q_1 = q_2 = 2.4 $
(hence $R_{\rm br}$ is dummy), which is a mean value for local
Seyfert~1 galaxies obtained by \citet{Patrick2012}, and
$R_{\rm out} = 10^3 \, r_{\rm G}$ (where $r_{\rm G}$ is the gravitational
radius of the SMBH). We fix $a=0$, which cannot be well constrained by the data.
The inclination $i$ is fixed at the value determined from the infrared
data by \citet{Ichikawa2015}. The photon index, normalization, and cutoff
energy are linked to those of the {\tt zcutoffpl} term.

\item The table models ({\tt atable\{xclumpy\_R.fits\}} and {\tt atable\{xclumpy\_L.fits\}}) correspond to the reflection continuum
    and emission lines from the torus, respectively, based on the
    XCLUMPY model. The parameters are the photon index,
    cutoff energy, equatorial hydrogen column density $(N^{\rm Equ}_{\rm
    H})$, torus angular width $( \sigma)$, and inclination angle
    $(i)$\footnote{The other parameters, the inner and outer radii of the torus, the radius of each clump,
    the number of clumps along the equatorial plane, and the index of
    radial density profile, are fixed at 0.05~pc, 1.00~pc 0.002~pc, 10.0 and 0.50,
    respectively \citep{Tanimoto2019a}.} . The photon index, normalization, and cutoff energy are linked
    to those of the {\tt zcutoffpl} term. The values of $\sigma$ and $i$ are
    fixed at those determined from the infrared data by \citet{Ichikawa2015}.

\item The {\tt zgauss} term represents an emission line
      feature at 0.77~keV in IC~4329A \citep{Brenneman2014}.

\end{enumerate}

We find that this model well reproduce the broadband spectra of both IC~4329A ($\chi^2$/dof = 1338.8/1120 ) and NGC~7469 ($\chi^2$/dof = 1668.7/1574). The
best-fit parameters are summarized in the second columns of Tables \ref{tab2-par} and
\ref{tab3-par}, and the best-fit models folded with the responses and the fitting
residuals are plotted in Figures~\ref{fig1-fitting}(a) and \ref{fig1-fitting}(c) for IC~4329A and NGC~7469,
respectively. The best-fit models in units of $E I(E)$, where $I$ is
the energy flux at the energy $E$, are plotted in Figures~\ref{fig2-models}(a) and \ref{fig2-models}(c)
for IC~4329A and NGC~7469, respectively.

\subsection{Model 2: Partial Covering Model}

As an alternative interpretation of X-ray spectra of Seyfert~1 galaxies
to relativistic reflection models, \citet{Miyakawa2012} proposed the ``variable
partial covering'' model and applied it for the spectral and timing
analysis of MCG--6--30--15 (see also \citealt{Miller2008} for an
earlier work). In this model, the broad iron-K emission line feature is
produced mainly by a deep iron K-edge structure due to the partial
absorber in the line of sight. \citet{Iso2016} applied the same model to the
\textit{Suzaku} data of 20 nearby AGNs and found that it can reproduce the
observations including time variability. Our Model~2, which is based on
the model in \citet{Iso2016}, is expressed in the XSPEC terminology as follows:

\begin{eqnarray}
    \textrm{Model 2} & = & \mathsf{const * phabs} \nonumber\\
    & * &\mathsf{(zxipcf * zxipcf * (zcutoffpl + compTT) } \nonumber\\
    & + &\mathsf{ atable\{xclumpy\_R.fits\} + atable\{xclumpy\_L.fits\}} \nonumber\\
    & + &\mathsf{ zgauss) }. \nonumber\\
\end{eqnarray}

\begin{enumerate}
\renewcommand{\labelenumi}{(\arabic{enumi})}
    \item Same as Model 1-(1)
    \item The {\tt zcutoffpl} term represents the direct component and the {\tt compTT} term the soft excess. Two layers
	  of absorption by ionized matter ({\tt zxipcf}) are multiplied,
	  one of which is a partial absorber with a large column density
	  ($N_{\rm H} \sim 1 \times 10^{24} \rm cm^{-2}$).
Note that \citet{Miyakawa2012} considered three warm absorbers for MCG--6--30--15,
	  whereas two are sufficient to explain our spectra of IC~4329A and NGC~7469.
    \item Same as Model 1-(3)

\item Same as Model 1-(4)

\end{enumerate}

We also find that Model~2 gives fairly good description of the broadband
spectra for both targets ($\chi^2/$dof = 1321.2/1120 for IC~4329A and 1683.6/1574 for
NGC~7469). Tables \ref{tab2-par} and \ref{tab3-par} (third columns) summarize the best-fit
parameters, Figures~\ref{fig1-fitting}(b) and \ref{fig1-fitting}(d) plot the best-fit folded models and
the residuals, and Figures~\ref{fig2-models}(b) and \ref{fig2-models}(d) plot the best-fit models in
units of $E I(E)$ for IC~4329A and NGC~7469, respectively.

\section{Discussion}
\label{sec5}

We have presented the results of our application of the X-ray clumpy
torus model (XCLUMPY) to the broadband (0.3--70~keV) spectra of two
seyfert~1 galaxies, IC~4329A and NGC~7469. This is the first work that
utilizes the XCLUMPY model for type-1 AGNs. We find that both Model~1
(relativistic reflection model) and Model~2 (partial covering model) are
able to reproduce the observed spectra almost equally well. This
confirms the degeneracy in interpreting the physical origins of the
X-ray spectra of type-1 AGNs just by utilizing the time averaged
spectroscopy. As mentioned in Section~1, we do not aim to favor or
disfavor of either of the two interpretations in this paper.
Below, we compare our results of the two models with previous
works that adopted similar models (\S~5.1). Then, we discuss the torus
properties of these AGNs by comparing our XCLUMPY results with the
infrared results (\S~5.2), which is the main purpose of our work.

\subsection{Comparison with Previous Studies}

We have estimated realistic contribution of the reflection component
from the torus in the broadband X-ray spectrum. The contribution to the
total flux in the 10--50~keV band is found to be 1.8\% (Model 1) and 4.5\%
(Model 2) for IC~4329A, and 11\% (Model 1) and 14\% (Model 2) for NGC~7469.
In IC~4329A, we obtain a smaller $N^{\rm Equ}_{\rm H}$ value (hence a weaker
intensity) with Model~1 than with Model~2. This is probably because a
part of the narrow emission-line flux can also be accounted for by
the emission line component in the RELXILL model under a limited energy
resolution. A future high energy resolution spectroscopy, like that by
{\it X-Ray Imaging and Spectroscopy Mission (XRISM)}
and {\it Athena}, would help us separate the two components. In the case
of NGC~7469, almost similar values are obtained with Models 1 and 2. In
NGC~7469, the best-fit RELXILL model produces a much broader Fe
K$\alpha$ line than that in the XCLUMPY model and hence such degeneracy
is ignorable.

Previous works often utilized either {\tt pexrav} or {\tt pexmon}
(\citealt{Gondoin2001}, \citealt{Iso2016}, and \citealt{Miyake2016} for IC~4329A;
\citealt{Nandra2000} and \citealt{Iso2016} for NGC~7469), or (unblurred) {\tt xillver} with $\xi=0$
(\citealt{Brenneman2014} for IC~4329A; \citealt{Middei2018} for NGC~7469)
to represent the reflection component from distant matter or to
approximate total reflection components including that from the
accretion disk. A notable difference is that XCLUMPY with $N^{\rm Equ}_{\rm H} \leq
 10^{24}~\rm cm^{-2}$ produces a much weaker reflection hump at $\sim$30~keV
 because the reflector is not as optically thick as assumed in
{\tt pexrav} or {\tt xillver}. In fact, when we replace XCLUMPY with
{\tt pexmon}\footnote{In the spectral fit the inclination is fixed at 60~degrees,
the photon index and cutoff energy are linked to those in the
direct component, and the reflection strength $\Omega/2\pi$ ($\Omega$ is
the solid angle of the reflector) is set free.}
in our models, the flux
of the torus reflection component at 30~keV is increased by a factor of 5.0 for IC~4329A
and by 2.0 for NGC~7469 compared with the case of XCLUMPY. It
inevitably affects estimates of the other spectral parameters. Thus, it
is very important to adopt a realistic model for the torus reflection for
correctly interpreting the broadband X-ray spectra of AGNs.

In Model~1 we consider two reflection components, one from the torus and
the other from the inner accretion disk utilizing the RELXILL code.
Our model yields $R_{\rm in} =87^{+73}_{-31}~r_{\rm G}$ for IC~4329A,
and $R_{\rm in} = 9.4^{+26}_{-3.4}~r_{\rm G}$ for NGC~7469,
by assuming the emissivity index $q=2.4$, the spin parameter $a=0$ and
the inclination angle ($i$) determined by the infrared data.
This would suggest that the accretion disk in IC~4329A
is truncated before reaching
the innermost stable orbit (ISCO), supporting the argument by
\citet{Done2000}.
By contrast, our result is consistent with the disk in NGC~7469
extending to the ISCO (i.e., $6~r_{\rm G}$ for $a=0$); this conclusion does not
change even if we adopt $a=0.998$ (maximum spin) instead.
\citet{Patrick2012} applied
the {\tt relline} model, the same code as in RELXILL,
to the \textit{Suzaku} and
\textit{Swift}/BAT spectra of nearby Seyfert~1s including our targets.
Although direct comparison is difficult because of parameter coupling
(they made $q$ and $i$ free but fixed either $a$ or $R_{\rm in}$),
their results are also consistent with a truncated disk in IC~4329A
($R_{\rm in}=37^{+8}_{-9}~r_{\rm G}$, $a=0.998$ (fixed), $q=2.3^{+0.3}_{-0.4}$, and $i=51^{+4}_{-3}$~degrees)
and with a disk extending to the ISCO in NGC~7469
($R_{in}=R_{\rm ISCO}$, $a=0.78^{+0.18}_{-0.17}$, $q=1.7^{+0.1}_{-0.8}$, and $i=80^{+8}_{-5}$~degrees, where $R_{\rm ISCO}$ is
the radius of the ISCO).

In Model~2, we confirm the claim by \citet{Iso2016} that a partial covering
model with a large column density of $N_{\rm H} \sim 10^{24}~\rm cm^{-2}$
can reproduce the data without invoking a relativistic reflection from
the accretion disk. The best-fit column density and ionization parameter
of the absorbers are not exactly the same as those in \citet{Iso2016}: they
obtained $N_{\rm H,2} = 1.62^{+0.27}_{-0.07} \times 10^{24}~\rm cm^{-2}$ and ${\rm log}\, \xi_2  = 1.49^{+0.23}_{-0.52}$ for IC~4329A and $N_{\rm H,2} = 1.64^{+1.06}_{-0.28} \times 10^{24}~\rm cm^{-2}$ and ${\rm log}\, \xi_2  = 1.42^{+0.58}_{-0.44}$ for NGC~7469 (their Table~2) and a full absorber (ZXIPCF1) is not required in
both targets. The discrepancy is probably because we utilized a
simplified {\tt zxipcf} model, whereas \citet{Iso2016} utilized their own XSTAR
based warm absorber model. Also, the difference between XCLUMPY and
{\tt pexrav}, which is utilized by \citet{Iso2016}, could affect the fit.
Application of more sophisticated absorption models utilizing XSTAR is
beyond the scope of this paper.

\subsection{Comparison with the Infrared Results}

In a type-1 AGN that shows no line-of-sight absorption, the equivalent
width of a narrow iron K$\alpha$ line
carries key information on the torus parameters
assuming that the contributions to the line flux from
the outer accretion disk and/or BLR are unimportant.
There is, however, degeneracy among the parameters, $N^{\rm Equ}_{\rm H}$, $\sigma$, and
$i$ (see Appendix). To avoid it, we have fixed $\sigma$ and $i$ at the
values obtained from the infrared observations by \citet{Ichikawa2015}.
This
enables us to make direct comparison on $N^{\rm Equ}_{\rm H}$ with the infrared
results. While X-rays measure all material (gas and dust) among which
gas is dominant in mass, infrared data are only sensitive to the amount
of dust. Thus, we can constrain the gas-to-dust ratio (in terms of the
ratio between the optical extinction and the X-ray column density) in
the torus.
In this analysis, it is implicitly assumed
that the gas and dust have the same spatial distribution.

\citet{Ichikawa2015} determined the V-band extinction along the
equatorial plane to be $A_{\rm V} = 1.93\pm0.17 \times 10^3~\rm mag$ for IC~4329A
and $A_{\rm V} = 1.75^{+0.30}_{-0.33} \times 10^3~\rm mag$ for NGC~7469.
Combining the X-ray results of $N^{\rm Equ}_{\rm H}$ with these $A_{\rm V}$ values,
we derive the $N_{\rm H}/A_{\rm V}$ ratios to be (2.8--7.4) $ \times 10^{19}~\rm{cm^{-2}~mag^{-1}}$ (IC~4329A)
and (4.7--8.2) $ \times 10^{20}~\rm{cm^{-2}~mag^{-1}}$ (NGC~7469),
which are smaller than the canonical value of Galactic ISM,
$1.87 \times 10^{21}~\rm{cm^{-2}~mag^{-1}}$ \citep{Draine2003},
by factors of 25--68 and 2.3--3.9 for IC~4329A and NGC~7469,
respectively; when we compare our results with those obtained by \citet{Maiolino2001},
the differences are further enhanced by factors of $\sim$ 3--100. This
result suggests that their tori are more ``dusty'' than, or have different dust
properties (e.g., size distribution) from Galactic ISM.

To confirm this trend in an alternative way,
we also perform spectral analysis by
fixing $N^{\rm Equ}_{\rm H}$ at the infrared results (converted with the
Galactic $N_{\rm H}/A_{\rm V}$ ratio) and leaving $\sigma$ free.
Then, we obtain
$\sigma \leq 10.1$~degrees (Model~1,2)\footnote{The lower limit is pegged at 10~degrees (the lower boundary
value in the table model).} for IC~4329A, and
$\sigma = 17.8^{+3.0}_{-6.2}$~degrees (Model~1) and $\sigma = 20.0^{+0.8}_{-3.0}$~degrees (Model~2) for NGC~7469.
These $\sigma$ values
are smaller than the infrared results,
$40\pm1$~degrees (IC~4329A) and $21\pm2$~degrees (NGC~7469).
In this picture, the gas-to-dust ratio is the same as
the Galactic ISM one at the equatorial plane
but rapidly decreases with the elevation angle
(i.e., the dust is vertically more extended than the gas),
making the ``averaged'' gas-to-dust ratio in the torus
smaller than the Galactic value.

On the basis of the X-ray results on $\sigma$,
it is possible to infer the covering fraction of the torus (with
$N_{\rm H} > 10^{22}~\rm cm^{-2}$), $C_{\rm T}$.
In XCLUMPY, the mean hydrogen column density at the elevation angle
$\theta \,(\equiv 90^\circ - i)$ is given by
\begin{eqnarray}
    N_{\rm H}\left(\theta\right) = N^{\rm Equ}_{\rm H} \exp\left(-\left(\frac{\theta}{\sigma}\right)^2\right).
\end{eqnarray}
Defining $\theta_{\rm c}$ such that $N_{\rm H}\left(\theta_{\rm c}\right) = 10^{22}~\rm cm^{-2}$,
we find $\theta_{\rm c} \leq 24.5$~degrees and $\theta_{\rm c}$ = 27.9--50.0~degrees,
which corresponding to $C_{\rm T} \leq 0.41$ and $C_{\rm T}$ = 0.47--0.77 for
IC~4329A and NGC~7469, respectively.
We can compare these values with the predictions from
\citet{Ricci2017Nature}, who showed that the Eddington ratio
$\lambda_{\rm Edd}$ is the key parameter that determines the torus geometry.
At $\lambda _{\rm Edd} =$ 0.13--0.16 (for IC~4329A) and
$\lambda _{\rm Edd} =$ 0.36--0.40 (for NGC~7469),
\footnote{Here we adopt the black hole masses $M_{\rm BH} =
1.2 \times 10^8~M_\odot$ for IC~4329A \citep{deLaCalle2010}
and
$M_{\rm BH} = 1.1 \times 10^7~M_\odot$ for NGC~7469 \citep{Peterson2014}, and
convert the 2--10~keV luminosities to bolometric ones with a correction
factor of 30.}
the mean torus covering fractions are predicted to be
$\sim$0.22--0.44 and $\sim$0.22--0.42, respectively \citep{Ricci2017Nature}.
Our result for IC~4329A is consistent with this prediction, whereas
that for NGC~7469 is larger than it.
We might be overestimating the torus reflection component in NGC~7469 because
of possible contributions from the outer accretion disk
and/or BLR in the observed iron-K line flux. Even if it is the case,
it strengthens our conclusion that the torus is dusty.

Our results suggest that at least a non-negligible fraction of AGNs have
more ``dusty'' torus than Galactic ISM.
This argument
may seem to be opposite to the
well-known previous results reporting larger $N_{\rm H}/A_{\rm V}$
values than in Galactic ISM in some AGNs (e.g., \citealt{Maiolino2001}).
However, we note that the sample of \citet{Maiolino2001} is not an unbiased
one but is consisting of AGNs that show cold absorption in X-rays and
optical broad emission lines. On the basis of the ``dust color'' method
to measure optical extinction, \citealt{Burtscher2016} showed that $N_{\rm
H}/A_{\rm V}$ values were on average consistent with the Galactic value
when variability in $N_{\rm H}$ due to dust-free gas in the BLR
is considered, except for heavily obscured AGNs.
In fact, by
applying the XCLUMPY model to 12 obscured AGNs in the \citet{Ichikawa2015}
sample,
Tanimoto et al. (in prep.) report that the mean value of
$N_{\rm H}/A_{\rm V}$ in the line-of-sight material
is similar to the Galactic ISM value, although
there is a large scatter ($\sim$1 dex) among the sample.
The reason behind the variation in
$N_{\rm H}/A_{\rm V}$ among individual objects is unclear.
The mid-infrared to X-ray luminosity ratios
of IC~4329A and NGC~7469 are located close to the standard correlation
\citep{Ichikawa2019}, apparently implying that the total amounts of dust
are not largely different from other AGNs.
Further systematic studies using a larger sample
would be required to solve this issue.

\section{Conclusion}
\label{sec6}

In this paper, we have reported the application of the X-ray clumpy
torus model XCLUMPY to the broadband spectra of two Seyfert~1 galaxies,
IC~4329A and NGC~7469, whose torus parameters were obtained from the
infrared spectra \citep{Ichikawa2015}. This is the first work that utilizes
XCLUMPY for unobscured AGNs. We have shown that the intensity of the
narrow Fe K$\alpha$ fluorescence line can be used to infer the torus
geometry, even in type-1 AGNs that show no line-of-sight absorption. The
main conclusions are summarized below.

\begin{enumerate}
    \item We are able to well produce the simultaneously observed
	  broadband (0.3--70~keV) spectra of both targets with two
	  different models containing XCLUMPY: (Model~1) a relativistic
	  reflection model and (Model~2) partial covering model. By
	  fixing the angular width and inclination at the values
	  determined by the infrared spectra, we constrain the column
	  density along the equatorial plane for each object.

    \item The XCLUMPY component (i.e., the torus reflection component)
	  produces a weaker hump structure at $\sim$30~keV compared
	  with other reflection models such as {\tt pexmon} and {\tt xillver}.
	  This fact must be correctly taken into account when interpreting
	  AGN broadband spectra.

     \item By comparing with the infrared results, the $N_{\rm H}/A_{\rm
	  V}$ ratios are found to be by factors of 25--68 and 2.3--3.9
	  smaller than the Galactic ISM value for IC~4329A and NGC~7469,
	  respectively. This is opposite to the trend reported for some
	  AGNs (e.g., \citealt{Maiolino2001}). Our results suggest that a
	  non-negligible fraction of AGNs have more ``dusty'' tori than
	  Galactic ISM.

\end{enumerate}

\acknowledgements

We thank the anonymous referee for a careful reading of our manuscript
and comments that helped us to improve the quality of the paper.
We thank Dr. Kohei Ichikawa for discussions.
Part of this work was financially supported by the Grant-in-Aid for
Scientific Research 17K05384 (Y.U.) and for JSPS fellows for young
researchers (A.T.). This research has made use of the \textit{NuSTAR}
Data Analysis Software (NUSTARDAS) jointly developed by the ASI Science
Data Center (ASDC, Italy and the California Institute of Technology
(Caltech, USA). This research also made use of data obtained with
\textit{XMM-Newton}, an ESA science mission with instruments and
contributions directly funded by ESA Member States and NASA, and use of
public \textit{Suzaku} data obtained through the Data ARchives and
Transmission System (DARTS) provided by the Institute of Space and
Astronautical Science (ISAS) at the Japan Aerospace Exploration Agency
(JAXA). For data reduction, we used software provided by the High Energy
Astrophysics Science Archive Research Center (HEASARC) at NASA/Goddard
Space Flight Center.

\facilities{\textit{XMM-Newton}, \textit{Suzaku}, \textit{NuSTAR}.}

\software{HEAsoft (HEASARC 2014), SAS (v17.0.0; \citealt{Gabriel2004}), NUSTARDAS, XSPEC \citep{Arnaud1996}, RELXILL \citep{Dauser2013,Garcia2014}, XCLUMPY \citep{Tanimoto2019a}.}

\begin{figure*}
\epsscale{1.1}
\plottwo{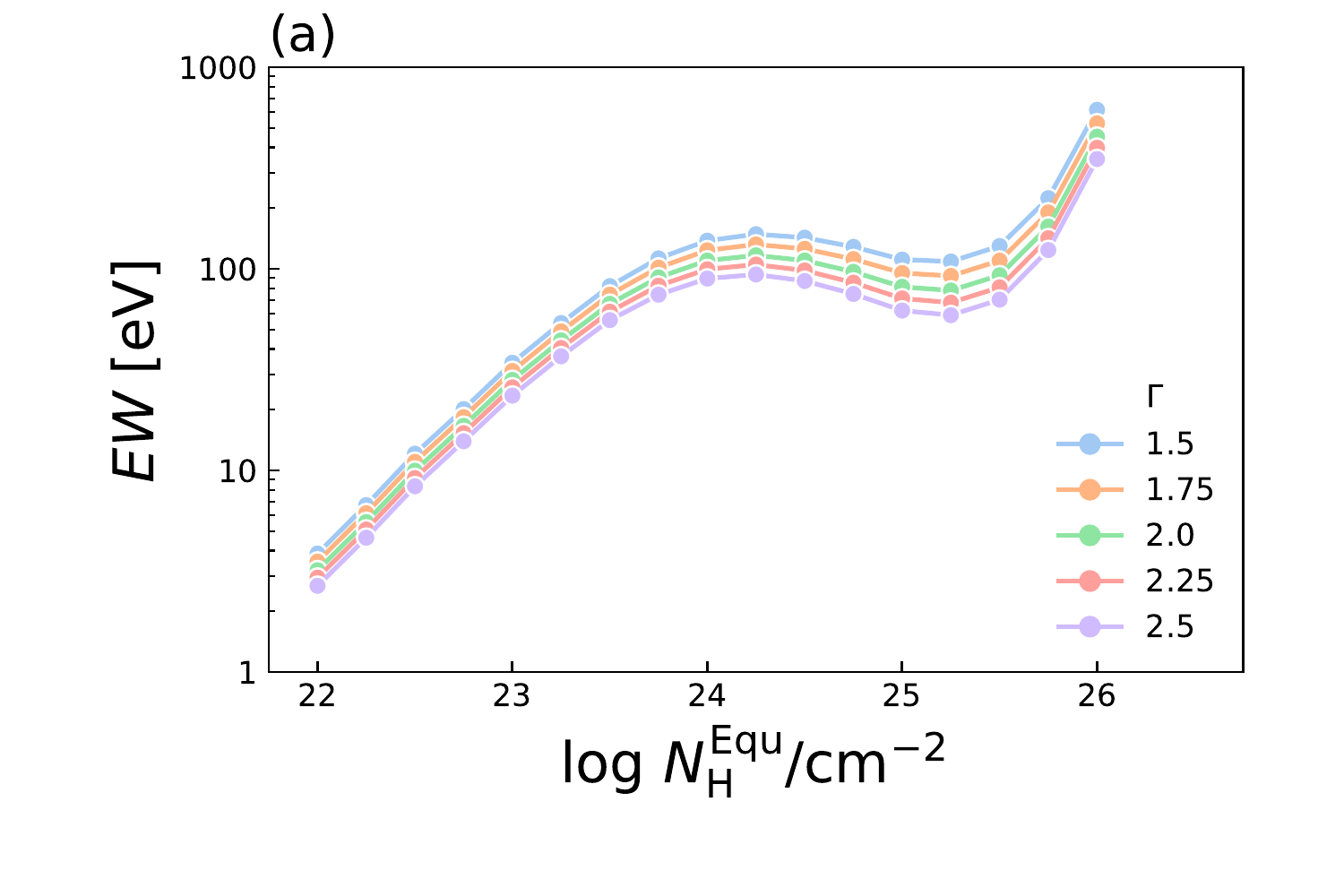}{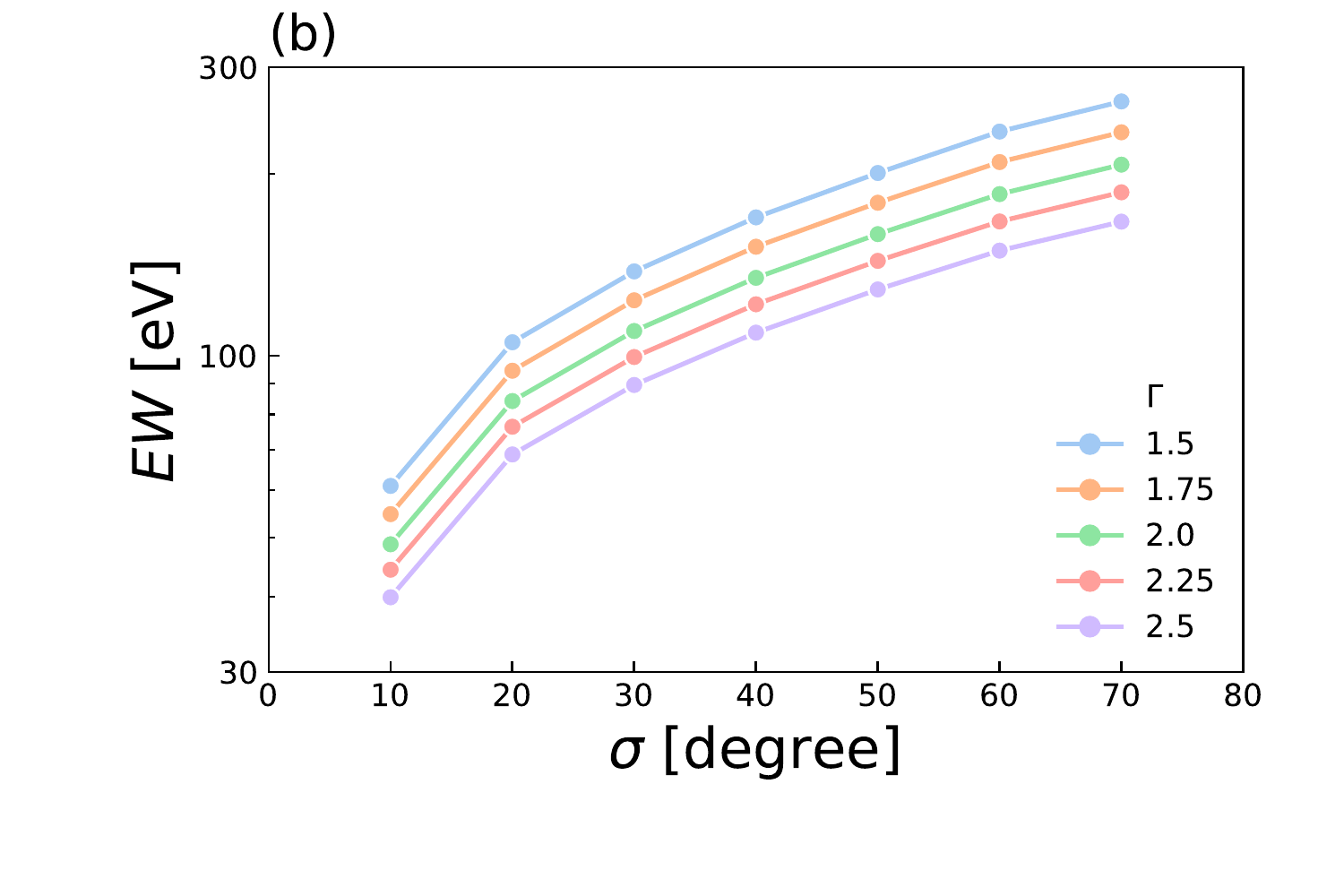}
\epsscale{0.55}
\plotone{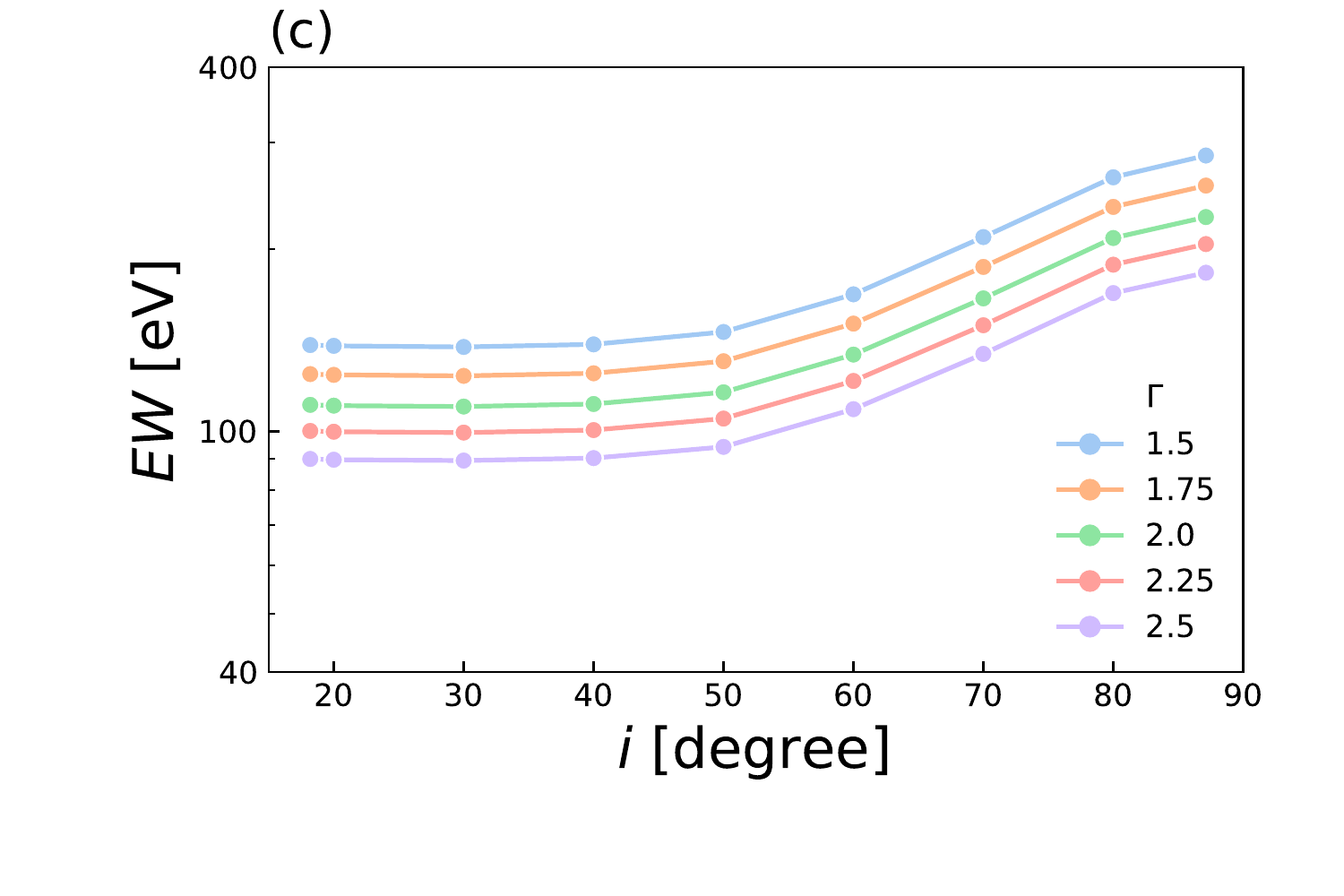}
\caption{Predicted iron-K equivalent width for type-1 AGNs plotted as a
function of torus parameter: (a) hydrogen column density along the
equatorial plane, (b) torus angular width, and (c) inclination angle.
Blue, orange, green, red, and violet lines correspond to
photon indices of $\Gamma =$ 1.5,
1.75, 2.0, 2.25, 2.5, respectively. We adopt $\log N^{\rm Equ}_{\rm H} /
\rm{cm^{-2}}= 24.0$, $\sigma = 30$~degrees, $i = 30$~degrees, and $E_{\rm
Cut} = 100$~keV.
We note that the line-of-sight absorption is $\log N_{\rm H} \sim 24$ in the case of $\log N^{\rm Equ}_{\rm H} \sim 26$.
} \label{fig3}
\end{figure*}

\appendix
\section{Predicted Fe K$\alpha$ Equivalent Width from XCLUMPY for Unobscured
AGNs}

Using XCLUMPY model, we investigate dependences of predicted iron-K
equivalent width on torus parameters: hydrogen column density along the
equatorial plane ($N^{\rm Equ}_{\rm H}$), torus angular width ($\sigma$), and
inclination angle ($i$). Figure~\ref{fig3} plots the predicted iron-K
equivalent width for type-1 AGNs as a function of the torus parameters.

\bibliographystyle{aasjournal}
\bibliography{reference}

\begin{thebibliography}{}
\expandafter\ifx\csname natexlab\endcsname\relax\def\natexlab#1{#1}\fi
\providecommand{\url}[1]{\href{#1}{#1}}

\bibitem[{{Anders} \& {Grevesse}(1989)}]{Anders&Grevesse1989}
{Anders}, E., \& {Grevesse}, N. 1989, Geochimica et Cosmochimica Acta, 53, 197

\bibitem[{{Arnaud}(1996)}]{Arnaud1996}
{Arnaud}, K.~A. 1996, in Astronomical Society of the Pacific Conference Series,
  Vol. 101, Astronomical Data Analysis Software and Systems V, ed. G.~H.
  {Jacoby} \& J.~{Barnes}, 17

\bibitem[{{Barcons} {et~al.}(2003){Barcons}, {Carrera}, \&
  {Ceballos}}]{Barcons2003}
{Barcons}, X., {Carrera}, F.~J., \& {Ceballos}, M.~T. 2003, \mnras, 339, 757

\bibitem[{{Beckmann} {et~al.}(2006){Beckmann}, {Gehrels}, {Shrader}, \&
  {Soldi}}]{Beckmann2006}
{Beckmann}, V., {Gehrels}, N., {Shrader}, C.~R., \& {Soldi}, S. 2006, \apj,
  638, 642

\bibitem[{{Blustin} {et~al.}(2003){Blustin}, {Branduardi-Raymont}, {Behar},
  {Kaastra}, {Kriss}, {Page}, {Kahn}, {Sako}, \& {Steenbrugge}}]{Blustin2003}
{Blustin}, A.~J., {Branduardi-Raymont}, G., {Behar}, E., {et~al.} 2003, \aap,
  403, 481

\bibitem[{{Brenneman} {et~al.}(2014){Brenneman}, {Madejski}, {Fuerst}, {Matt},
  {Elvis}, {Harrison}, {Ballantyne}, {Boggs}, {Christensen}, {Craig}, {Fabian},
  {Grefenstette}, {Hailey}, {Madsen}, {Marinucci}, {Rivers}, {Stern}, {Walton},
  \& {Zhang}}]{Brenneman2014}
{Brenneman}, L.~W., {Madejski}, G., {Fuerst}, F., {et~al.} 2014, \apj, 788, 61

\bibitem[{{Burtscher} {et~al.}(2016){Burtscher}, {Davies}, {Graci{\'a}-Carpio},
  {Koss}, {Lin}, {Lutz}, {Nandra}, {Netzer}, {Orban de Xivry}, {Ricci},
  {Rosario}, {Veilleux}, {Contursi}, {Genzel}, {Schnorr-M{\"u}ller},
  {Sternberg}, {Sturm}, \& {Tacconi}}]{Burtscher2016}
{Burtscher}, L., {Davies}, R.~I., {Graci{\'a}-Carpio}, J., {et~al.} 2016, \aap,
  586, A28

\bibitem[{{Dauser} {et~al.}(2013){Dauser}, {Garcia}, {Wilms}, {B{\"o}ck},
  {Brenneman}, {Falanga}, {Fukumura}, \& {Reynolds}}]{Dauser2013}
{Dauser}, T., {Garcia}, J., {Wilms}, J., {et~al.} 2013, \mnras, 430, 1694

\bibitem[{{Dauser} {et~al.}(2010){Dauser}, {Wilms}, {Reynolds}, \&
  {Brenneman}}]{Dauser2010}
{Dauser}, T., {Wilms}, J., {Reynolds}, C.~S., \& {Brenneman}, L.~W. 2010,
  \mnras, 409, 1534

\bibitem[{{de La Calle P{\'e}rez} {et~al.}(2010){de La Calle P{\'e}rez},
  {Longinotti}, {Guainazzi}, {Bianchi}, {Dov{\v{c}}iak}, {Cappi}, {Matt},
  {Miniutti}, {Petrucci}, {Piconcelli}, {Ponti}, {Porquet}, \& {Santos-
  Lle{\'o}}}]{deLaCalle2010}
{de La Calle P{\'e}rez}, I., {Longinotti}, A.~L., {Guainazzi}, M., {et~al.}
  2010, \aap, 524, A50

\bibitem[{{De Rosa} {et~al.}(2002){De Rosa}, {Fabian}, \& {Piro}}]{Rosa2002}
{De Rosa}, A., {Fabian}, A.~C., \& {Piro}, L. 2002, \mnras, 334, L21

\bibitem[{{Done} {et~al.}(2000){Done}, {Madejski}, \& {{\.Z}ycki}}]{Done2000}
{Done}, C., {Madejski}, G.~M., \& {{\.Z}ycki}, P.~T. 2000, \apj, 536, 213

\bibitem[{{Draine}(2003)}]{Draine2003}
{Draine}, B.~T. 2003, Annual Review of Astronomy and Astrophysics, 41, 241

\bibitem[{{Fukazawa} {et~al.}(2009){Fukazawa}, {Mizuno}, {Watanabe}, {Kokubun},
  {Takahashi}, {Kawano}, {Nishino}, {Sasada}, {Shirai}, {Takahashi}, {Umeki},
  {Yamasaki}, {Yasuda}, {Bamba}, {Ohno}, {Takahashi}, {Ushio}, {Enoto},
  {Kitaguchi}, {Makishima}, {Nakazawa}, {Uehara}, {Yamada}, {Yuasa}, {Isobe},
  {Kawaharada}, {Tanaka}, {Tashiro}, {Terada}, \& {Yamaoka}}]{Fukazawa2009}
{Fukazawa}, Y., {Mizuno}, T., {Watanabe}, S., {et~al.} 2009, Publications of
  the Astronomical Society of Japan, 61, S17

\bibitem[{{Gabriel} {et~al.}(2004){Gabriel}, {Denby}, {Fyfe}, {Hoar}, {Ibarra},
  {Ojero}, {Osborne}, {Saxton}, {Lammers}, \& {Vacanti}}]{Gabriel2004}
{Gabriel}, C., {Denby}, M., {Fyfe}, D.~J., {et~al.} 2004, in Astronomical
  Society of the Pacific Conference Series, Vol. 314, Astronomical Data
  Analysis Software and Systems (ADASS) XIII, ed. F.~{Ochsenbein}, M.~G.
  {Allen}, \& D.~{Egret}, 759

\bibitem[{{Garc{\'\i}a} {et~al.}(2014){Garc{\'\i}a}, {Dauser}, {Lohfink},
  {Kallman}, {Steiner}, {McClintock}, {Brenneman}, {Wilms}, {Eikmann},
  {Reynolds}, \& {Tombesi}}]{Garcia2014}
{Garc{\'\i}a}, J., {Dauser}, T., {Lohfink}, A., {et~al.} 2014, \apj, 782, 76

\bibitem[{{Gondoin} {et~al.}(2001){Gondoin}, {Barr}, {Lumb}, {Oosterbroek},
  {Orr}, \& {Parmar}}]{Gondoin2001}
{Gondoin}, P., {Barr}, P., {Lumb}, D., {et~al.} 2001, \aap, 378, 806

\bibitem[{{Gonz{\'a}lez-Mart{\'\i}n} {et~al.}(2013){Gonz{\'a}lez-Mart{\'\i}n},
  {Rodr{\'\i}guez-Espinosa}, {D{\'\i}az-Santos}, {Packham}, {Alonso-Herrero},
  {Esquej}, {Ramos Almeida}, {Mason}, \& {Telesco}}]{Gonzalez2013}
{Gonz{\'a}lez-Mart{\'\i}n}, O., {Rodr{\'\i}guez-Espinosa}, J.~M.,
  {D{\'\i}az-Santos}, T., {et~al.} 2013, \aap, 553, A35

\bibitem[{{Guainazzi} {et~al.}(1994){Guainazzi}, {Matsuoka}, {Piro}, {Mihara},
  \& {Yamauchi}}]{Guainazzi1994}
{Guainazzi}, M., {Matsuoka}, M., {Piro}, L., {Mihara}, T., \& {Yamauchi}, M.
  1994, \apj, 436, L35

\bibitem[{{Harrison} {et~al.}(2013){Harrison}, {Craig}, {Christensen},
  {Hailey}, {Zhang}, {Boggs}, {Stern}, {Cook}, {Forster}, {Giommi},
  {Grefenstette}, {Kim}, {Kitaguchi}, {Koglin}, {Madsen}, {Mao}, {Miyasaka},
  {Mori}, {Perri}, {Pivovaroff}, {Puccetti}, {Rana}, {Westergaard}, {Willis},
  {Zoglauer}, {An}, {Bachetti}, {Barri{\`e}re}, {Bellm}, {Bhalerao},
  {Brejnholt}, {Fuerst}, {Liebe}, {Markwardt}, {Nynka}, {Vogel}, {Walton},
  {Wik}, {Alexander}, {Cominsky}, {Hornschemeier}, {Hornstrup}, {Kaspi},
  {Madejski}, {Matt}, {Molendi}, {Smith}, {Tomsick}, {Ajello}, {Ballantyne},
  {Balokovi{\'c}}, {Barret}, {Bauer}, {Blandford}, {Brandt}, {Brenneman},
  {Chiang}, {Chakrabarty}, {Chenevez}, {Comastri}, {Dufour}, {Elvis}, {Fabian},
  {Farrah}, {Fryer}, {Gotthelf}, {Grindlay}, {Helfand}, {Krivonos}, {Meier},
  {Miller}, {Natalucci}, {Ogle}, {Ofek}, {Ptak}, {Reynolds}, {Rigby},
  {Tagliaferri}, {Thorsett}, {Treister}, \& {Urry}}]{Harrison2013}
{Harrison}, F.~A., {Craig}, W.~W., {Christensen}, F.~E., {et~al.} 2013, \apj,
  770, 103

\bibitem[{{Huang} {et~al.}(2011){Huang}, {Wang}, {Tan}, {Yang}, \&
  {Huang}}]{Huang2011}
{Huang}, X.-X., {Wang}, J.-X., {Tan}, Y., {Yang}, H., \& {Huang}, Y.-F. 2011,
  \apj, 734, L16

\bibitem[{{Ichikawa} {et~al.}(2015){Ichikawa}, {Packham}, {Ramos Almeida},
  {Asensio Ramos}, {Alonso-Herrero}, {Gonz{\'a}lez-Mart{\'\i}n}, {Lopez-
  Rodriguez}, {Ueda}, {D{\'\i}az- Santos}, {Elitzur}, {H{\"o}nig}, {Imanishi},
  {Levenson}, {Mason}, {Perlman}, \& {Alsip}}]{Ichikawa2015}
{Ichikawa}, K., {Packham}, C., {Ramos Almeida}, C., {et~al.} 2015, \apj, 803,
  57

\bibitem[{{Ichikawa} {et~al.}(2019){Ichikawa}, {Ricci}, {Ueda}, {Bauer},
  {Kawamuro}, {Koss}, {Oh}, {Rosario}, {Shimizu}, {Stalevski}, {Fuller},
  {Packham}, \& {Trakhtenbrot}}]{Ichikawa2019}
{Ichikawa}, K., {Ricci}, C., {Ueda}, Y., {et~al.} 2019, \apj, 870, 31

\bibitem[{{Ishisaki} {et~al.}(2007){Ishisaki}, {Maeda}, {Fujimoto}, {Ozaki},
  {Ebisawa}, {Takahashi}, {Ueda}, {Ogasaka}, {Ptak}, {Mukai}, {Hamaguchi},
  {Hirayama}, {Kotani}, {Kubo}, {Shibata}, {Ebara}, {Furuzawa}, {Iizuka},
  {Inoue}, {Mori}, {Okada}, {Yokoyama}, {Matsumoto}, {Nakajima}, {Yamaguchi},
  {Anabuki}, {Tawa}, {Nagai}, {Katsuda}, {Hayashida}, {Bamba}, {Miller},
  {Sato}, \& {Yamasaki}}]{Ishisaki2007}
{Ishisaki}, Y., {Maeda}, Y., {Fujimoto}, R., {et~al.} 2007, Publications of the
  Astronomical Society of Japan, 59, 113

\bibitem[{{Iso} {et~al.}(2016){Iso}, {Ebisawa}, {Sameshima}, {Mizumoto},
  {Miyakawa}, {Inoue}, \& {Yamasaki}}]{Iso2016}
{Iso}, N., {Ebisawa}, K., {Sameshima}, H., {et~al.} 2016, Publications of the
  Astronomical Society of Japan, 68, S27

\bibitem[{{Jansen} {et~al.}(2001){Jansen}, {Lumb}, {Altieri}, {Clavel}, {Ehle},
  {Erd}, {Gabriel}, {Guainazzi}, {Gondoin}, {Much}, {Munoz}, {Santos},
  {Schartel}, {Texier}, \& {Vacanti}}]{Jansen2001}
{Jansen}, F., {Lumb}, D., {Altieri}, B., {et~al.} 2001, \aap, 365, L1

\bibitem[{{Kalberla} {et~al.}(2005){Kalberla}, {Burton}, {Hartmann}, {Arnal},
  {Bajaja}, {Morras}, \& {P{\"o}ppel}}]{Kalberla2005}
{Kalberla}, P.~M.~W., {Burton}, W.~B., {Hartmann}, D., {et~al.} 2005, \aap,
  440, 775

\bibitem[{{Kawaguchi} {et~al.}(2001){Kawaguchi}, {Shimura}, \&
  {Mineshige}}]{Kawaguchi2001}
{Kawaguchi}, T., {Shimura}, T., \& {Mineshige}, S. 2001, \apj, 546, 966

\bibitem[{{Kawamuro} {et~al.}(2016){Kawamuro}, {Ueda}, {Tazaki}, {Terashima},
  \& {Mushotzky}}]{Kawamuro2016LLAGN}
{Kawamuro}, T., {Ueda}, Y., {Tazaki}, F., {Terashima}, Y., \& {Mushotzky}, R.
  2016, \apj, 831, 37

\bibitem[{{Kokubun} {et~al.}(2007){Kokubun}, {Makishima}, {Takahashi},
  {Murakami}, {Tashiro}, {Fukazawa}, {Kamae}, {Madejski}, {Nakazawa},
  {Yamaoka}, {Terada}, {Yonetoku}, {Watanabe}, {Tamagawa}, {Mizuno}, {Kubota},
  {Isobe}, {Takahashi}, {Sato}, {Takahashi}, {Hong}, {Kawaharada}, {Kawano},
  {Mitani}, {Murashima}, {Suzuki}, {Abe}, {Miyawaki}, {Ohno}, {Tanaka},
  {Yanagida}, {Itoh}, {Ohnuki}, {Tamura}, {Endo}, {Hirakuri}, {Hiruta},
  {Kitaguchi}, {Kishishita}, {Sugita}, {Takahashi}, {Takeda}, {Enoto},
  {Hirasawa}, {Katsuta}, {Matsumura}, {Onda}, {Sato}, {Ushio}, {Ishikawa},
  {Murase}, {Odaka}, {Suzuki}, {Yaji}, {Yamada}, {Yamasaki}, {Yuasa}, \& {Hxd
  Team}}]{Kokubun2007}
{Kokubun}, M., {Makishima}, K., {Takahashi}, T., {et~al.} 2007, Publications of
  the Astronomical Society of Japan, 59, 53

\bibitem[{{Koyama} {et~al.}(2007){Koyama}, {Tsunemi}, {Dotani}, {Bautz},
  {Hayashida}, {Tsuru}, {Matsumoto}, {Ogawara}, {Ricker}, {Doty}, {Kissel},
  {Foster}, {Nakajima}, {Yamaguchi}, {Mori}, {Sakano}, {Hamaguchi},
  {Nishiuchi}, {Miyata}, {Torii}, {Namiki}, {Katsuda}, {Matsuura}, {Miyauchi},
  {Anabuki}, {Tawa}, {Ozaki}, {Murakami}, {Maeda}, {Ichikawa}, {Prigozhin},
  {Boughan}, {Lamarr}, {Miller}, {Burke}, {Gregory}, {Pillsbury}, {Bamba},
  {Hiraga}, {Senda}, {Katayama}, {Kitamoto}, {Tsujimoto}, {Kohmura}, {Tsuboi},
  \& {Awaki}}]{Koyama2007}
{Koyama}, K., {Tsunemi}, H., {Dotani}, T., {et~al.} 2007, Publications of the
  Astronomical Society of Japan, 59, 23

\bibitem[{{Magdziarz} \& {Zdziarski}(1995)}]{Magdziarz1995}
{Magdziarz}, P., \& {Zdziarski}, A.~A. 1995, \mnras, 273, 837

\bibitem[{{Maiolino} {et~al.}(2001){Maiolino}, {Marconi}, {Salvati},
  {Risaliti}, {Severgnini}, {Oliva}, {La Franca}, \& {Vanzi}}]{Maiolino2001}
{Maiolino}, R., {Marconi}, A., {Salvati}, M., {et~al.} 2001, \aap, 365, 28

\bibitem[{{McKernan} \& {Yaqoob}(2004)}]{McKernan2004}
{McKernan}, B., \& {Yaqoob}, T. 2004, \apj, 608, 157

\bibitem[{{Middei} {et~al.}(2018){Middei}, {Bianchi}, {Cappi}, {Petrucci},
  {Ursini}, {Arav}, {Behar}, {Branduardi- Raymont}, {Costantini}, {De Marco},
  {Di Gesu}, {Ebrero}, {Kaastra}, {Kaspi}, {Kriss}, {Mao}, {Mehdipour},
  {Paltani}, {Peretz}, \& {Ponti}}]{Middei2018}
{Middei}, R., {Bianchi}, S., {Cappi}, M., {et~al.} 2018, \aap, 615, A163

\bibitem[{{Miller} {et~al.}(2008){Miller}, {Turner}, \& {Reeves}}]{Miller2008}
{Miller}, L., {Turner}, T.~J., \& {Reeves}, J.~N. 2008, \aap, 483, 437

\bibitem[{{Mitsuda} {et~al.}(2007){Mitsuda}, {Bautz}, {Inoue}, {Kelley},
  {Koyama}, {Kunieda}, {Makishima}, {Ogawara}, {Petre}, {Takahashi}, {Tsunemi},
  {White}, {Anabuki}, {Angelini}, {Arnaud}, {Awaki}, {Bamba}, {Boyce}, {Brown},
  {Chan}, {Cottam}, {Dotani}, {Doty}, {Ebisawa}, {Ezoe}, {Fabian}, {Figueroa},
  {Fujimoto}, {Fukazawa}, {Furusho}, {Furuzawa}, {Gendreau}, {Griffiths},
  {Haba}, {Hamaguchi}, {Harrus}, {Hasinger}, {Hatsukade}, {Hayashida}, {Henry},
  {Hiraga}, {Holt}, {Hornschemeier}, {Hughes}, {Hwang}, {Ishida}, {Ishisaki},
  {Isobe}, {Itoh}, {Iyomoto}, {Kahn}, {Kamae}, {Katagiri}, {Kataoka},
  {Katayama}, {Kawai}, {Kilbourne}, {Kinugasa}, {Kissel}, {Kitamoto}, {Kohama},
  {Kohmura}, {Kokubun}, {Kotani}, {Kotoku}, {Kubota}, {Madejski}, {Maeda},
  {Makino}, {Markowitz}, {Matsumoto}, {Matsumoto}, {Matsuoka}, {Matsushita},
  {McCammon}, {Mihara}, {Misaki}, {Miyata}, {Mizuno}, {Mori}, {Mori}, {Morii},
  {Moseley}, {Mukai}, {Murakami}, {Murakami}, {Mushotzky}, {Nagase}, {Namiki},
  {Negoro}, {Nakazawa}, {Nousek}, {Okajima}, {Ogasaka}, {Ohashi}, {Oshima},
  {Ota}, {Ozaki}, {Ozawa}, {Parmar}, {Pence}, {Porter}, {Reeves}, {Ricker},
  {Sakurai}, {Sanders}, {Senda}, {Serlemitsos}, {Shibata}, {Soong}, {Smith},
  {Suzuki}, {Szymkowiak}, {Takahashi}, {Tamagawa}, {Tamura}, {Tamura},
  {Tanaka}, {Tashiro}, {Tawara}, {Terada}, {Terashima}, {Tomida}, {Torii},
  {Tsuboi}, {Tsujimoto}, {Tsuru}, {Turner}, {Ueda}, {Ueno}, {Ueno}, {Uno},
  {Urata}, {Watanabe}, {Yamamoto}, {Yamaoka}, {Yamasaki}, {Yamashita},
  {Yamauchi}, {Yamauchi}, {Yaqoob}, {Yonetoku}, \& {Yoshida}}]{Mitsuda2007}
{Mitsuda}, K., {Bautz}, M., {Inoue}, H., {et~al.} 2007, Publications of the
  Astronomical Society of Japan, 59, S1

\bibitem[{{Miyakawa} {et~al.}(2012){Miyakawa}, {Ebisawa}, \&
  {Inoue}}]{Miyakawa2012}
{Miyakawa}, T., {Ebisawa}, K., \& {Inoue}, H. 2012, Publications of the
  Astronomical Society of Japan, 64, 140

\bibitem[{{Miyake} {et~al.}(2016){Miyake}, {Noda}, {Yamada}, {Makishima}, \&
  {Nakazawa}}]{Miyake2016}
{Miyake}, K., {Noda}, H., {Yamada}, S., {Makishima}, K., \& {Nakazawa}, K.
  2016, Publications of the Astronomical Society of Japan, 68, S28

\bibitem[{{Nandra} {et~al.}(2000){Nandra}, {Le}, {George}, {Edelson},
  {Mushotzky}, {Peterson}, \& {Turner}}]{Nandra2000}
{Nandra}, K., {Le}, T., {George}, I.~M., {et~al.} 2000, \apj, 544, 734

\bibitem[{{Nandra} {et~al.}(2007){Nandra}, {O'Neill}, {George}, \&
  {Reeves}}]{Nandra2007}
{Nandra}, K., {O'Neill}, P.~M., {George}, I.~M., \& {Reeves}, J.~N. 2007,
  \mnras, 382, 194

\bibitem[{{Nenkova} {et~al.}(2008{\natexlab{a}}){Nenkova}, {Sirocky},
  {Ivezi{\'c}}, \& {Elitzur}}]{Nenkova2008a}
{Nenkova}, M., {Sirocky}, M.~M., {Ivezi{\'c}}, {\v{Z}}., \& {Elitzur}, M.
  2008{\natexlab{a}}, \apj, 685, 147

\bibitem[{{Nenkova} {et~al.}(2008{\natexlab{b}}){Nenkova}, {Sirocky},
  {Nikutta}, {Ivezi{\'c}}, \& {Elitzur}}]{Nenkova2008b}
{Nenkova}, M., {Sirocky}, M.~M., {Nikutta}, R., {Ivezi{\'c}}, {\v{Z}}., \&
  {Elitzur}, M. 2008{\natexlab{b}}, \apj, 685, 160

\bibitem[{{Noda} {et~al.}(2013){Noda}, {Makishima}, {Nakazawa}, \&
  {Yamada}}]{Noda2013a}
{Noda}, H., {Makishima}, K., {Nakazawa}, K., \& {Yamada}, S. 2013, \apj, 771,
  100

\bibitem[{{Odaka} {et~al.}(2016){Odaka}, {Yoneda}, {Takahashi}, \&
  {Fabian}}]{Odaka2016}
{Odaka}, H., {Yoneda}, H., {Takahashi}, T., \& {Fabian}, A. 2016, \mnras, 462,
  2366

\bibitem[{{Ordov{\'a}s-Pascual} {et~al.}(2017){Ordov{\'a}s-Pascual}, {Mateos},
  {Carrera}, {Wiersema}, {Barcons}, {Braito}, {Caccianiga}, {Del Moro}, {Della
  Ceca}, \& {Severgnini}}]{Ordovas-Pascual2017}
{Ordov{\'a}s-Pascual}, I., {Mateos}, S., {Carrera}, F.~J., {et~al.} 2017,
  \mnras, 469, 693

\bibitem[{{Patrick} {et~al.}(2012){Patrick}, {Reeves}, {Porquet}, {Markowitz},
  {Braito}, \& {Lobban}}]{Patrick2012}
{Patrick}, A.~R., {Reeves}, J.~N., {Porquet}, D., {et~al.} 2012, \mnras, 426,
  2522

\bibitem[{{Peterson} {et~al.}(2014){Peterson}, {Grier}, {Horne}, {Pogge},
  {Bentz}, {De Rosa}, {Denney}, {Martini}, {Sergeev}, {Kaspi}, {Minezaki},
  {Zu}, {Kochanek}, {Siverd}, {Shappee}, {Araya Salvo}, {Beatty}, {Bird},
  {Bord}, {Borman}, {Che}, {Chen}, {Cohen}, {Dietrich}, {Doroshenko}, {Drake},
  {Efimov}, {Free}, {Ginsburg}, {Henderson}, {King}, {Koshida}, {Mogren},
  {Molina}, {Mosquera}, {Motohara}, {Nazarov}, {Okhmat}, {Pejcha}, {Rafter},
  {Shields}, {Skowron}, {Skowron}, {Valluri}, {van Saders}, \&
  {Yoshii}}]{Peterson2014}
{Peterson}, B.~M., {Grier}, C.~J., {Horne}, K., {et~al.} 2014, \apj, 795, 149

\bibitem[{{Ramos Almeida} \& {Ricci}(2017)}]{RamosAlmeidaRicci2017}
{Ramos Almeida}, C., \& {Ricci}, C. 2017, Nature Astronomy, 1, 679

\bibitem[{{Ricci} {et~al.}(2014){Ricci}, {Ueda}, {Ichikawa}, {Paltani},
  {Boissay}, {Gandhi}, {Stalevski}, \& {Awaki}}]{Ricci2014A&A}
{Ricci}, C., {Ueda}, Y., {Ichikawa}, K., {et~al.} 2014, \aap, 567, A142

\bibitem[{{Ricci} {et~al.}(2017){Ricci}, {Trakhtenbrot}, {Koss}, {Ueda},
  {Schawinski}, {Oh}, {Lamperti}, {Mushotzky}, {Treister}, {Ho}, {Weigel},
  {Bauer}, {Paltani}, {Fabian}, {Xie}, \& {Gehrels}}]{Ricci2017Nature}
{Ricci}, C., {Trakhtenbrot}, B., {Koss}, M.~J., {et~al.} 2017, \nat, 549, 488

\bibitem[{{Risaliti} \& {Elvis}(2004)}]{Risaliti2004}
{Risaliti}, G., \& {Elvis}, M. 2004, in Astrophysics and Space Science Library,
  Vol. 308, Supermassive Black Holes in the Distant Universe, ed. A.~J.
  {Barger}, 187

\bibitem[{{Risaliti} {et~al.}(2005){Risaliti}, {Elvis}, {Fabbiano}, {Baldi}, \&
  {Zezas}}]{Risaliti2005}
{Risaliti}, G., {Elvis}, M., {Fabbiano}, G., {Baldi}, A., \& {Zezas}, A. 2005,
  \apj, 623, L93

\bibitem[{{Scott} {et~al.}(2005){Scott}, {Kriss}, {Lee}, {Quijano},
  {Brotherton}, {Canizares}, {Green}, {Hutchings}, {Kaiser}, {Marshall},
  {Oegerle}, {Ogle}, \& {Zheng}}]{Scott2005}
{Scott}, J.~E., {Kriss}, G.~A., {Lee}, J.~C., {et~al.} 2005, \apj, 634, 193

\bibitem[{{Shu} {et~al.}(2010){Shu}, {Yaqoob}, \& {Wang}}]{Shu2010}
{Shu}, X.~W., {Yaqoob}, T., \& {Wang}, J.~X. 2010, The Astrophysical Journal
  Supplement Series, 187, 581

\bibitem[{{Springob} {et~al.}(2005){Springob}, {Haynes}, {Giovanelli}, \&
  {Kent}}]{Springob2005}
{Springob}, C.~M., {Haynes}, M.~P., {Giovanelli}, R., \& {Kent}, B.~R. 2005,
  The Astrophysical Journal Supplement Series, 160, 149

\bibitem[{{Steenbrugge} {et~al.}(2005){Steenbrugge}, {Kaastra}, {Sako},
  {Branduardi-Raymont}, {Behar}, {Paerels}, {Blustin}, \&
  {Kahn}}]{Steenbrugge2005}
{Steenbrugge}, K.~C., {Kaastra}, J.~S., {Sako}, M., {et~al.} 2005, \aap, 432,
  453

\bibitem[{{Str{\"u}der} {et~al.}(2001){Str{\"u}der}, {Briel}, {Dennerl},
  {Hartmann}, {Kendziorra}, {Meidinger}, {Pfeffermann}, {Reppin}, {Aschenbach},
  {Bornemann}, {Br{\"a}uninger}, {Burkert}, {Elender}, {Freyberg}, {Haberl},
  {Hartner}, {Heuschmann}, {Hippmann}, {Kastelic}, {Kemmer}, {Kettenring},
  {Kink}, {Krause}, {M{\"u}ller}, {Oppitz}, {Pietsch}, {Popp}, {Predehl},
  {Read}, {Stephan}, {St{\"o}tter}, {Tr{\"u}mper}, {Holl}, {Kemmer}, {Soltau},
  {St{\"o}tter}, {Weber}, {Weichert}, {von Zanthier}, {Carathanassis}, {Lutz},
  {Richter}, {Solc}, {B{\"o}ttcher}, {Kuster}, {Staubert}, {Abbey}, {Holland},
  {Turner}, {Balasini}, {Bignami}, {La Palombara}, {Villa}, {Buttler},
  {Gianini}, {Lain{\'e}}, {Lumb}, \& {Dhez}}]{Struder2001}
{Str{\"u}der}, L., {Briel}, U., {Dennerl}, K., {et~al.} 2001, \aap, 365, L18

\bibitem[{{Takahashi} {et~al.}(2007){Takahashi}, {Abe}, {Endo}, {Endo}, {Ezoe},
  {Fukazawa}, {Hamaya}, {Hirakuri}, {Hong}, {Horii}, {Inoue}, {Isobe}, {Itoh},
  {Iyomoto}, {Kamae}, {Kasama}, {Kataoka}, {Kato}, {Kawaharada}, {Kawano},
  {Kawashima}, {Kawasoe}, {Kishishita}, {Kitaguchi}, {Kobayashi}, {Kokubun},
  {Kotoku}, {Kouda}, {Kubota}, {Kuroda}, {Madejski}, {Makishima}, {Masukawa},
  {Matsumoto}, {Mitani}, {Miyawaki}, {Mizuno}, {Mori}, {Mori}, {Murashima},
  {Murakami}, {Nakazawa}, {Niko}, {Nomachi}, {Okada}, {Ohno}, {Oonuki}, {Ota},
  {Ozawa}, {Sato}, {Shinoda}, {Sugiho}, {Suzuki}, {Taguchi}, {Takahashi},
  {Takahashi}, {Takeda}, {Tamura}, {Tamura}, {Tanaka}, {Tanihata}, {Tashiro},
  {Terada}, {Tominaga}, {Uchiyama}, {Watanabe}, {Yamaoka}, {Yanagida}, \&
  {Yonetoku}}]{Takahashi2007}
{Takahashi}, T., {Abe}, K., {Endo}, M., {et~al.} 2007, Publications of the
  Astronomical Society of Japan, 59, 35

\bibitem[{{Tanaka} {et~al.}(1995){Tanaka}, {Nandra}, {Fabian}, {Inoue},
  {Otani}, {Dotani}, {Hayashida}, {Iwasawa}, {Kii}, {Kunieda}, {Makino}, \&
  {Matsuoka}}]{Tanaka1995}
{Tanaka}, Y., {Nandra}, K., {Fabian}, A.~C., {et~al.} 1995, \nat, 375, 659

\bibitem[{{Tanimoto} {et~al.}(2019){Tanimoto}, {Ueda}, {Odaka}, {Kawaguchi},
  {Fukazawa}, \& {Kawamuro}}]{Tanimoto2019a}
{Tanimoto}, A., {Ueda}, Y., {Odaka}, H., {et~al.} 2019, \apj

\bibitem[{{Tazaki} {et~al.}(2013){Tazaki}, {Ueda}, {Terashima}, {Mushotzky}, \&
  {Tombesi}}]{Tazaki2013}
{Tazaki}, F., {Ueda}, Y., {Terashima}, Y., {Mushotzky}, R.~F., \& {Tombesi}, F.
  2013, \apj, 772, 38

\bibitem[{{Titarchuk}(1994)}]{Titarchuk1994}
{Titarchuk}, L. 1994, \apj, 434, 570

\bibitem[{{Turner} {et~al.}(2001){Turner}, {Abbey}, {Arnaud}, {Balasini},
  {Barbera}, {Belsole}, {Bennie}, {Bernard}, {Bignami}, {Boer}, {Briel},
  {Butler}, {Cara}, {Chabaud}, {Cole}, {Collura}, {Conte}, {Cros}, {Denby},
  {Dhez}, {Di Coco}, {Dowson}, {Ferrando}, {Ghizzardi}, {Gianotti}, {Goodall},
  {Gretton}, {Griffiths}, {Hainaut}, {Hochedez}, {Holland}, {Jourdain},
  {Kendziorra}, {Lagostina}, {Laine}, {La Palombara}, {Lortholary}, {Lumb},
  {Marty}, {Molendi}, {Pigot}, {Poindron}, {Pounds}, {Reeves}, {Reppin},
  {Rothenflug}, {Salvetat}, {Sauvageot}, {Schmitt}, {Sembay}, {Short},
  {Spragg}, {Stephen}, {Str{\"u}der}, {Tiengo}, {Trifoglio}, {Tr{\"u}mper},
  {Vercellone}, {Vigroux}, {Villa}, {Ward}, {Whitehead}, \&
  {Zonca}}]{Turner2001}
{Turner}, M.~J.~L., {Abbey}, A., {Arnaud}, M., {et~al.} 2001, \aap, 365, L27

\bibitem[{{Vasudevan} {et~al.}(2009){Vasudevan}, {Mushotzky}, {Winter}, \&
  {Fabian}}]{Vasudevan2009}
{Vasudevan}, R.~V., {Mushotzky}, R.~F., {Winter}, L.~M., \& {Fabian}, A.~C.
  2009, \mnras, 399, 1553

\bibitem[{{Willmer} {et~al.}(1991){Willmer}, {Focardi}, {Chan}, {Pellegrini},
  \& {da Costa}}]{Willmer1991}
{Willmer}, C.~N.~A., {Focardi}, P., {Chan}, R., {Pellegrini}, P.~S., \& {da
  Costa}, N.~L. 1991, \aj, 101, 57

\bibitem[{{Winter} {et~al.}(2009){Winter}, {Mushotzky}, {Reynolds}, \&
  {Tueller}}]{Winter2009}
{Winter}, L.~M., {Mushotzky}, R.~F., {Reynolds}, C.~S., \& {Tueller}, J. 2009,
  \apj, 690, 1322

\bibitem[{{Yaqoob} \& {Warwick}(1991)}]{Yaqoob1991}
{Yaqoob}, T., \& {Warwick}, R.~S. 1991, \mnras, 248, 773

\end{thebibliography}

\end{document}